\documentclass{emulateapj}
\usepackage{apjfonts}
\usepackage{epsfig}
\slugcomment{Accepted to ApJ}
\shorttitle{Radiative Acceleration of MHD Winds}
\shortauthors{Everett}

\begin{document}

\title{Radiative Transfer and Acceleration in Magnetocentrifugal Winds}

\author{John E. Everett} 
\affil{Canadian Institute of Theoretical Astrophysics, University of
Toronto, 60 Saint George Street, Toronto, ON M5S 3H8, Canada}
\email{everett@cita.utoronto.ca} 

\begin{abstract}
Detailed photoionization and radiative acceleration of self-similar
magnetocentrifugal accretion disk winds are explored.  First, a
general-purpose hybrid magnetocentrifugal and radiatively-driven wind
model is defined.  Solutions are then examined to determine how
radiative acceleration modifies magnetocentrifugal winds and how those
winds can influence radiative driving in Active Galactic Nuclei
(AGNs).  For the models studied here, both radiative acceleration by
bound-free (``continuum-driving'') and bound-bound (``line-driving'')
processes are found to be important, although magnetic driving
dominates the mass outflow rate for the Eddington ratios studied
($L/L_{\rm Edd} = 0.001 - 0.1$).  The solutions show that shielding
by a magnetocentrifugal wind can increase the efficiency of a
radiatively-driven wind, and also that, within a magnetocentrifugal
wind, radiative acceleration is sensitive to both the column in the
shield, the column of the wind and the initial density at the base of
the wind.
\end{abstract}

\section{Introduction}\label{Intro}

A variety of observational signatures point to the importance of
outflowing gas within many types of Active Galactic Nuclei (AGNs).
Blueshifted absorption features \citep[in Broad Absorption Line
Quasars, or BALQSOs; see, e.g.,][]{Weymann91} are seen in
approximately 15\% \citep{Reichard03} of radio-quiet quasars, with
velocities up to 0.1c. In addition, radio-loud quasars display
relativistic, collimated outflows.  More recently, both UV and X-ray
absorbing gas have been observed in approximately 60\% of Seyfert 1
galaxies \citep{C99}.  Observational estimates hint that the mass
outflow rate is nearly equal to the mass inflow rate \citep[for a
review of mass outflow in AGNs, see][]{CKG03}.

The development of models to explain the mass outflow rates, geometry,
and general kinematics of these winds has proven difficult, but
progress on a number of possibilities has been encouraging.  From very
early on, researchers examined both radiative wind models
\citep[e.g.,][]{DB84, VS88} and, to explain radio jets, hydromagnetic
models \citep[e.g.,][hereafter BP82]{BP82}.  Later models of
radiatively-driven winds were able to explain the BALQSO outflow
velocities and population fraction \citep[][hereafter, MCGV95]{MCGV95}
as well as the single-peaked emission lines \citep{MC97}.  The
line-driven models were also demonstrated in hydrodynamical
simulations \citep{PSD98, PSK00, Pereyra04}, which elucidated the
density structure, geometry, and kinematics of the flow in two
dimensions.  In addition, models with combinations of continuum- and
line-driving by X-rays were presented by \citet{CN01}.  One of the
concerns with line-driving in AGNs has been the possible
over-ionization of the gas by X-rays in the central continuum (leaving
the wind with too few lines to intercept flux in the lines): the need
for ``shielding gas'' to prevent this overionization has been a
persistent concern \citep[e.g., MCGV95,][]{CN03b, PK04}.

In addition, hydromagnetic winds have also been developed to gain
insight into the observations; some of these models have included
radiative driving.  In these magnetocentrifugal winds, gas is loaded
onto and tied to large-scale magnetic field lines; those field lines
then fling the matter centrifugally away from the disk, like beads
along a wire.  Magnetocentrifugal acceleration is commonly used to
explain large-scale collimated outflows in radio galaxies \citep[BP82;
for reviews of magnetocentrifugal driving, see][]{S96,KP00,F03}.  In
the context of the Broad Emission Line Region (BLR),
magnetocentrifugal outflows made up of clouds were first called upon
to explain the single-peaked broad emission lines, outflow velocities,
and densities \citep[][]{EBS92} .  The ``torus'' \citep[obscuring gas
that plays a central role in the Unification model, yielding a
dependence of observed properties with inclination angle;
see][]{Ant93} has been explained as a dusty, continuous magnetic wind
\citep[][hereafter KK94]{KK94} where radiative acceleration on dust
affects the wind geometry. Hydromagnetic disk winds with radiation
pressure were also developed by \citet{dKB95} as an alternative
explanation for the population fraction of BALQSOs and to understand
cloud confinement within the outflow.  In addition, another
magnetocentrifugal wind model was developed to explain single peaked
emission lines arising from a distribution of clouds \citep{Bot97}, as
well as the dynamics of the warm absorber in NGC 5548 \citep{BKS00}.
Finally, some effects of magnetic fields (not including
magnetocentrifugal driving as in BP82) have also been considered in
two-dimensional hydrodynamic simulations \citep{P00, P03}.


At the present time, both radiatively driven winds and
magnetocentrifugally driven winds (with radiative acceleration added
in some models) offer compelling but competing pictures of the key
physics in the cores of AGNs. No clear observational evidence yet
discriminates the dominant physics of wind-launching in AGNs.  For
radiatively-driven winds, three main lines of evidence hint that
radiative driving should be important in models of wind
dynamics. First, \citet{LB02} find a correlation between UV luminosity
and the observed outflow velocity, which agrees with a basic
prediction of line-driven wind models \citep{P99}.  In addition, the
``ghost of Ly-$\alpha$'' \citep{Arav95, CN03b}, as well as the
realization that the radiative momentum removed from the continuum in
blueshifted absorption lines in BALQSOs is a significant fraction of
the total radiative momentum (MCGV95) are both also important clues
that radiative acceleration is an important component to any
self-consistent model of AGN winds.  There are still, however,
concerns about how the ``shielding'' of the wind works \citep{CN03b,
PK04}.  In addition, in the case of stellar disk winds, observations
of two nova-like variables seem to show that their winds are not
dominated by radiative driving \citep{Hartley02}.  \citep[This
conclusion, however, rests on the prediction that line equivalent
widths are direct measures of mass outflow rate, which may not be the
case; see][]{Pereyra04}

On the other hand, the leading model for collimated radio jets in AGNs
(BP82) already calls upon large scale, dynamically important magnetic
fields, as do the hydromagnetic wind models \citep[mentioned above,
e.g.][]{KK94,K95,Bot97} that have also had success in explaining AGN
observations.  In addition, such a hydromagnetic wind would have no
difficulty with overionization, and so might naturally serve as a
``shield'' for radiative acceleration further from the central source.
There are, however, currently no models which address the interplay
between magnetocentrifugal driving and radiative acceleration; if such
wind models could be constrained, we may be able to observationally
distinguish the physics of wind launching in the cores of AGNs, and
gain insight into the role that outflows play both in accretion and in
feedback of those winds in the galaxy and surrounding matter.



This paper develops a detailed photoionization and dynamical model for
magnetocentrifugal winds in AGNs.  The model is designed to explore
the radiative transfer within such magnetocentrifugal winds, but also
to help understand how radiative driving impacts the kinematic
structure of such winds.  Constructing such a model builds the
foundation for later work to determine absorption and emission line
profiles in order to compare with observations and check for the
presence of magnetocentrifugal winds within AGNs.  An overview of this
model is presented in \S\ref{modelOverview}; the model is then defined
in detail in \S\ref{detailedModel}.  An examination of the structure
of a particular ``fiducial'' magnetocentrifugal,
radiatively-accelerated wind is described in \S\ref{fiducialModel} and
then the dependences of radiative acceleration on some initial
parameters are shown in \S\ref{modelDependences}.  Conclusions and
directions for future work are summarized in \S\ref{results}.

\section{Model Overview}\label{modelOverview}

\begin{figure*}[ht]
\begin{center}
\includegraphics[width=12cm]{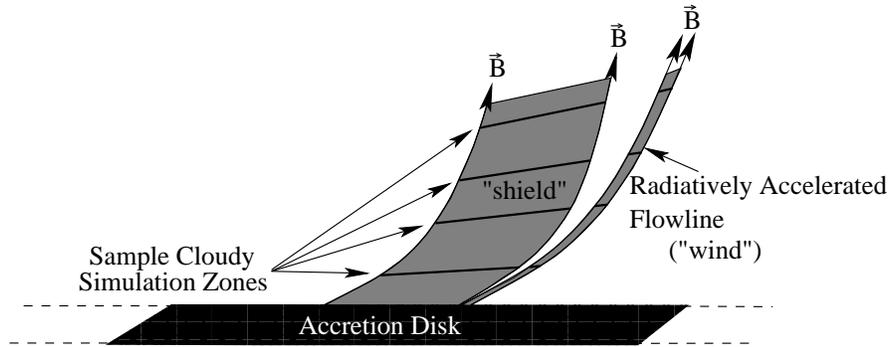}
\caption{Schematic of the basic geometry and major components of the
radiatively accelerated magnetocentrifugal wind model.  The wind
itself is split into two components: a pure magnetocentrifugal wind
that acts as a shield (the wider component on the left) and an
optically thin streamline with combined magnetocentrifugal and
radiative acceleration.  The heavy black lines indicate a few of the
radial zones where Cloudy simulates the photoionization balance of the
wind (typically, $\sim$~40 such radial Cloudy simulations are run,
spaced logarithmically in polar angle, $\theta$).
\label{thinWind}}
\end{center}
\end{figure*}

Before examining the model in detail, it is instructive to present a
summary of the basic design.  The semianalytic model developed here
includes magnetic acceleration and radiative acceleration of a
continuous wind launched from an accretion disk.  A detailed treatment
of radiative transfer is included by using Version 96.00 of the
photoionization simulation program Cloudy, last described by
\citet{F98}.  These elements are introduced in an approximate
schematic of the wind model shown in Figure~\ref{thinWind}, depicting
a portion of the accretion disk and outflowing wind.  In this figure,
radiation (entering from the left side of the schematic) first
encounters a purely magnetocentrifugally accelerated wind, which will
be referred to as the ``shield'', as it absorbs radiation from the
central continuum.  The shield is introduced as a separate component
in order to cleanly differentiate the effect of shielding from
radiative acceleration; radiative driving of the shield is therefore
not considered in this work.  Beyond that shield is an optically thin,
radiatively and magnetically accelerated wind streamline (which we
will henceforth refer to as the ``wind'').  In this portion of the
model, both magnetocentrifugal and radiative forces help accelerate
the flow off of the accretion disk; the magnetic fieldlines are shown
by the black lines bordering the outflow.  The included radiative
acceleration is calculated by first simulating the photoionization
within both the shield and the wind along radial paths such as the
thick, black lines in the figure.
 
The separation of the wind into two components is, of course,
artificial.  In reality, radiative acceleration would gradually
increase in importance for portions of the wind that are increasingly
shielded.  However, splitting the outflow into these two components
allows a first-order, qualitative solution that can be used to gain
some understanding of how magnetic and radiative forces might
interact, and how a magnetic wind may be able to act as a radiative
``shield'' to allow more efficient radiative acceleration.  While
artificial, this method of using two wind components has already been
used successfully to examine winds with magnetocentrifugal and
radiative driving on dust \citep[e.g., KK94,][]{K95}.


       
\begin{figure*}[ht]
\includegraphics[width=18cm]{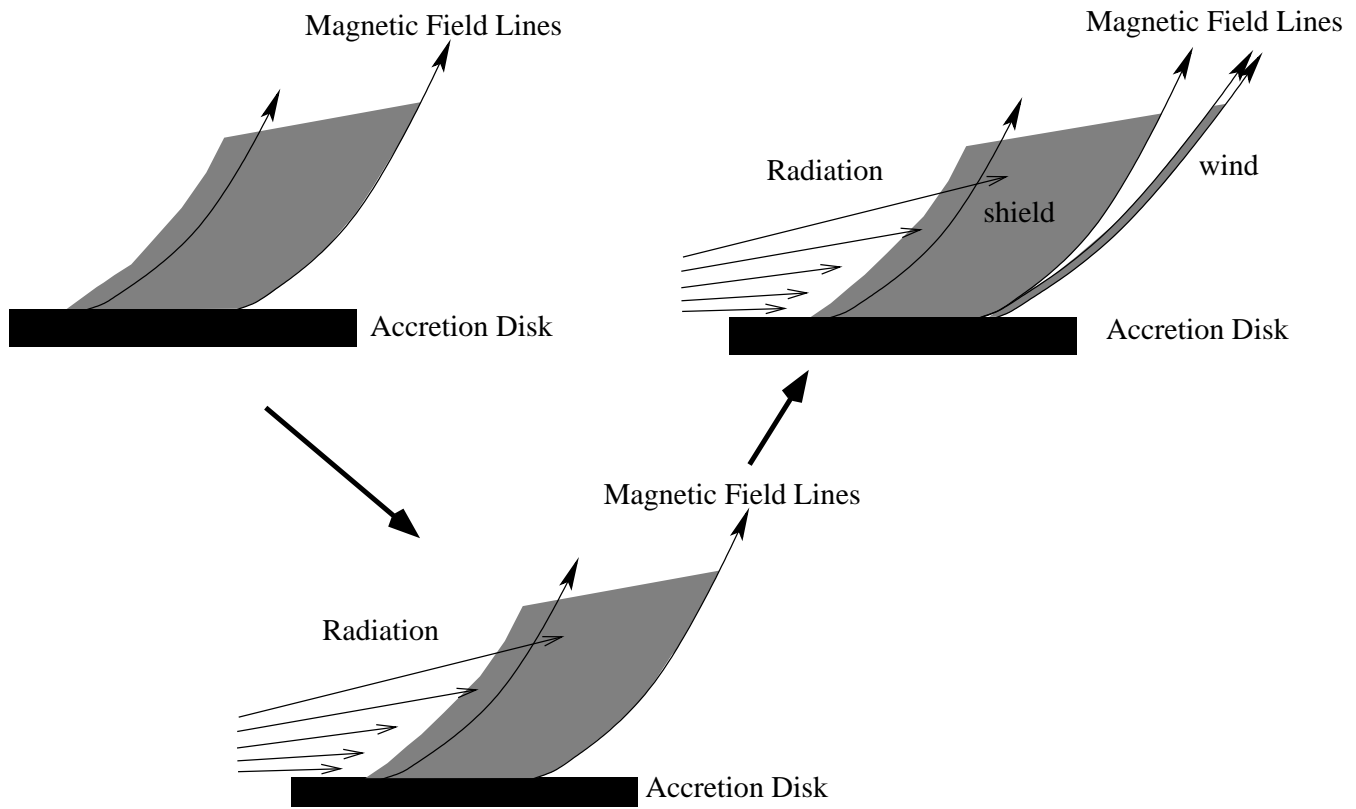} 
\caption{The iterative scheme in the wind model, showing a single
iteration of the model, moving counterclockwise around the figure.
The model starts by solving the self-similar magnetohydrodynamic (MHD)
equations for the structure of a centrifugally driven outflow ({\it upper
left}).  Next, photoionization simulations determine the continuum
incident on the radiatively accelerated streamline as well as the
ionization state of the gas in the streamline, and then those results
are utilized to calculate the radiative acceleration of that flow.
The radiative acceleration is then applied (as a function of the polar
angle $\theta$) to the second component of the self-similar
magnetocentrifugal wind.  A new wind structure for that second section
of the outflow is then derived from the interaction of those forces
({\it upper right}).
\label{flowchart}}
\end{figure*}

Figure~\ref{flowchart} presents a schematic flow chart of how model
calculations proceed.  The wind starts as a self-similar
magnetohydrodynamic model that yields the pure magnetocentrifugal wind
solution (covered in \S\ref{mhdWind}).  Next, simulations of the
photoionization balance of that wind streamline (\S\ref{photoSims})
are run, and the resultant ionization balance and transmitted
continuum are used to calculate the radiative acceleration behind the
shield (\S\ref{radAccCalc}).  Next, the radiative acceleration is
input (as a function of polar angle, $\theta$) back into the
self-similar magnetohydrodynamic model, modifying the structure of the
wind streamline, while leaving the shield unaffected.  This process is
then repeated, simulating the photoionization of that modified wind
and recalculating the radiative acceleration terms.  We typically
iterate five to eight times to converge to a final equilibrium
solution.


With the basic model now summarized, it may be instructive to compare
and contrast it to the recent wind model examined in \citet{P03},
where the combination of magnetic and radiative forces in disk winds
is also investigated.  \citet{P03} concentrates on numerical
simulations of time-dependent winds with line driving and magnetic
forces.  These numerical simulations allow large-scale models of
outflows that are valuable in understanding global wind structures in
many different astrophysical contexts.  In contrast, the
radiatively-driven components of this model are more localized, since
radiative acceleration is applied to the shielded streamline
only. This model setup has been chosen for its flexibility in
radiative acceleration modeling; using Cloudy for such radiative
acceleration models enables computations not only of radiative line
driving but also of continuum driving, and allows us the freedom to
include dust and easily vary the incident spectrum, for example.  In
addition, the magnetic winds produced in \citet{P03} are not
magnetocentrifugal outflows (as in BP82), as are the semianalytic
winds presented here.  Further, these new models are steady-state, not
time-dependent.  Finally, semi-analytic, steady-state models allow an
exploration of general behaviors through many parameter variations;
large-scale numerical simulations can usually vary only a few
parameters.  In summary, these models cover different facets of the
disk-wind problem, yielding different perspectives on a complicated
system.

\section{The Two-Phase Hydromagnetic and Radiative Wind Model}\label{detailedModel}

In this section, we describe in detail the model's components, and
derive key equations.  The magnetocentrifugal wind model is introduced first
(\S\ref{mhdWind}), followed by the photoionization simulations
(\S\ref{photoSims}), radiative acceleration calculation
(\S\ref{radAccCalc}), and finally a discussion of the wind model
equation of motion (\S\ref{windEqMtn}).

\subsection{Magnetocentrifugal Self-Similar Wind Solution}\label{mhdWind}

To derive the equations governing the continuous magnetocentrifugal
wind, we start with the equations of a stationary, axisymmetric
magnetohydrodynamic (MHD) flow in cylindrical coordinates, make the
assumption of self-similarity in the spherical radial coordinate, and
utilize the continuity equation, conservation of angular momentum
along the flow, and both the radial and vertical momentum equations,
very much as in BP82 and KK94 (see Appendix~\ref{selfSimWindEqns}).
Thermal effects are neglected in the wind, therefore effectively
assuming that the wind starts out supersonic (this assumption is
checked later on, see \S\ref{assumptions}).  In deriving the wind
equations, we use the same simplifications as BP82, except for the
added complication that energy is not conserved in the
radiatively-accelerated system due to the constant input of radiative
energy into the wind.  In the original formulation of BP82,
conservation of energy supplied an additional constraint which allowed
a simplification of the equations of motion to two first-order
differential equations.  Because energy is not conserved in this flow,
the equivalent of three first-order differential equations must be
integrated, solving for three parameters simultaneously instead of two
as in BP82.  The detailed setup and derivation of this set of
equations of motion are given in Appendix~\ref{selfSimWindEqns}.

The integration of the momentum equations starts by specifying the
following initial parameters: the mass loading of the wind (the ratio
of mass flux to magnetic flux in the magnetocentrifugal wind, $\kappa
\propto \frac{4 \pi \rho v_p}{B_p}$, where $\rho$ is the mass density
of the wind, $v_p$ is the poloidal velocity of the wind
($v_p~\equiv~(v_r^2 + v_z^2)^{1/2}$), and $B_p$ is the poloidal
magnetic field strength), the specific angular momentum of gas and
field in the wind, and the power-law exponent ($b$) that describes the
change in density with spherical radius: $\rho \propto R^{-b}$.  Also
input, as parameters, are the mass of the central black hole,
$M_{\bullet}$, the wind's launch radius on the disk, $r_0$, and the
density at the base of the wind at the launch radius, $n_0$.  The
program employs a ``shooting'' algorithm \citep[using the SLATEC
routine DNSQ;][]{Powell70} to integrate from the singular point (the
Alfv\'en point) to the disk, solving for the height of the singular
point above the disk ($\chi_A$) and the slope of the streamline at
both the disk and the Alfv\'en point ($\xi'_0$ and $\xi'_A$) by
matching the integration results to boundary conditions on the disk.
After solving for the position of the Alfv\'en point, the equations of
motion are integrated from the disk to a user-specified height beyond
the Alfv\'en point; along the streamline, the run of velocity,
density, and magnetic field are calculated.

This code has been tested (without radiative acceleration) against the
solution given in BP82 and have duplicated their results to within
8\%.  This is close to the previously reported 4\% variance in the
recalculation of BP82 reported in \citet{S93}.  The difference in
these new results comes not only from using higher precision
calculations compared to BP82 \citep[we use the same precision
as][]{S93}, but in addition, a more complex set of equations is being
solved.

\subsection{Photoionization Simulations of the Wind}\label{photoSims}

Next, Version 96.00 of the photoionization code Cloudy \citep[][]{F98}
is used to simulate the absorption of the magnetocentrifugal shield
and wind as well as the ionization state at the wind streamline.
Photoionization simulations of the shield and wind are used to
calculate the radiative acceleration and allow considerable
flexibility in gas parameters (such as gas abundances, dust, central
continuum, etc).  This flexibility is gained at the cost of simulating
the photoionization state of the wind as if it were a static medium,
as Cloudy assumes; this is only true of the recombination timescale
for the gas is much shorter than the transit timescale of gas in the
region simulated ($\tau_{\rm recomb} < \tau_{\rm flow}$).  We have
however, verified, \textit{a posteriori}, that $\tau_{\rm recomb} <
\tau_{\rm flow}$ for all of the radiatively accelerated ions at the
base of the wind where radiative acceleration is important, using the
recombination rate approximations in \citet{AR92} and \citet{VF96} to
calculate the recombination times.  Even at high latitude, $\tau_{\rm
recomb} < \tau_{\rm flow}$ for all of the significant ions.  At the
highest latitudes, highly ionized O, which has the longest
recombination time among the significantly radiatively accelerated
ions, has a recombination timescale that is still a factor of two less
than the transit time.  Meanwhile, highly ionized Fe ions, which
dominate the low level of radiative acceleration at high latitudes,
have recombination timescales an order of magnitude less than the
transit timescale.  Using Cloudy is therefore a reasonable
approximation for the radiative equilibrium within our winds
(especially since, with our code, adiabatic and advective effects are
added, see \S\ref{contWindPhoto}).  As Cloudy enables a flexible,
self-consistent calculation of radiative acceleration, we accept this
approximation to enable these calculations.


Of basic importance to the photoionization simulations is the
illuminating spectral energy distribution (SED).  For the purposes of
these simulations, an SED adapted from \citet{RE04} is used (an
example of the SED is shown in Figure~\ref{transmittedCont}).  This
SED is input into Cloudy via the generic ``AGN'' continuum with
$T_{\rm blackbody} = 1.5 \times 10^5$~K, $\alpha_{\rm ox} = -1.43$
\citep{Elvis02}, $\alpha_{\rm UV} = -0.44$, and $\alpha_X = 0.9$
($\alpha_{\rm ox}$ defines a single power-law that would describe the
continuum between $2500$~\AA\ and 2~keV, $\alpha_{\rm UV}$ is the
slope of the low-energy component of the Big Blue Bump, and
$\alpha_{\rm X}$ is the X-ray power-law exponent; our value of
$\alpha_{\rm ox}$ is taken from the middle of the range 0.8 to 1.0
given in \citet{RE04}).



Spectral signatures and radiative acceleration also depend, of course,
on the column density in the wind.  Observations yield only rough
constraints for this, so these columns as left as free parameters; the
effect of varying these columns will be investigated in this paper.
The columns throughout the shield and wind are set by the columns at
the base of the shield and wind, denoted $N_{\rm H,0}$ for the hydrogen
column at the base of the wind.  As the wind rises above the disk and
accelerates, that column density ($N_{\rm H}$) drops as a function of
height due to mass conservation (an example of this is shown later in
Fig.~\ref{fiducialWind}).  Investigating the shielding ability of such
a dynamic shielding column is of central interest to this paper, and
will be addressed in \S\ref{shieldColumnVariations}.

\subsubsection{Photoionization of the Continuous Wind}\label{contWindPhoto}

As depicted in Figure~\ref{thinWind}, Cloudy simulates the
photoionization of the wind along radial sight lines through the
shield and through the wind, ending at the site of radiative
acceleration on the wind streamline. The continuum incident on the
wind streamline is also calculated. The photoionization state and
continuum at the end of the Cloudy calculations are recorded, and then
used to compute the radiative acceleration of that gas.  Finally, the
acceleration is tabulated and applied as a function of $\theta$ along
the wind streamline by inputing the angle-dependent radiative
acceleration into the gravitational term (this is covered in more
detail in Appendix A, see eqn. \ref{gammaInSelfSim}).


Whereas Cloudy is designed to simulate the photoionization balance of
gases as in the shield and wind, it cannot easily incorporate
adiabatic and advection effects: Cloudy has no knowledge of the
particular velocity profile of the wind in the overarching model, or
the temperature difference between successive photoionization models
as the wind climbs above the disk. Therefore, this model calculates
both advective heating and adiabatic cooling in the wind, and adds
those terms manually into Cloudy's simulations.  Both terms largely
cancel in the wind, and have only a negligible effect on outflow
dynamics, but they are included in all of the models for completeness.

\subsection{Radiative Acceleration Calculations}\label{radAccCalc}

The model then incorporates the above-mentioned results for the
ionization structure and radiation field to calculate the radiative
forces felt by the wind.  There are two different kinds of radiative
acceleration to consider: continuum acceleration (including radiative
acceleration on dust) and line acceleration.  It is convenient to
express the radiative acceleration in terms of $\Gamma(\theta)$,
\begin{eqnarray}
\Gamma(\theta) & \equiv & \frac{a_{\rm radiative}(\theta)}{g} \label{gammaDef}, 
\end{eqnarray}
where $a_{\rm radiative}$ is the acceleration due to radiation, and
$g$ is the local gravitational acceleration.  

\subsubsection{Line and Continuum Acceleration}\label{lineContAccel}

In general, for continuum and line acceleration, the radiative
acceleration is given by
\begin{eqnarray}
\Gamma & = &  \frac{\frac{n_e \sigma_T F}{\rho c} ( M_{\rm cont} +
M_{\rm lines})}{\frac{G M_{\bullet}}{r^2 + z^2}},
\end{eqnarray}
where $F$ is the local flux (the flux transmitted through both the
shield and wind column), $n_e$ is the electron density, $\rho$ is the
gas density, $c$ is the speed of light, $G$ is the gravitational
constant, $M_{\bullet}$ is the mass of the central black hole, $r$ and
$z$ are cylindrical coordinates centered on the black hole, and
$M_{\rm cont}$ \& $M_{\rm lines}$ are the ``force multipliers'' that
relate how much the radiative forces on the gas (on the line and
continuum opacity, respectively) exceed the radiative forces on
electrons alone.  They are given below in terms of the continuum
opacity, $\chi_{\nu}$, and the line opacity, $\chi_l$, for the
continuum and lines, respectively:
\begin{eqnarray}
M_{\rm cont} & = & \frac{1}{n_e \sigma_T F} \int \chi_{\nu} F_{\nu} d\nu,
\\
M_{\rm lines} & = & \frac{1}{F} \sum_l F_l \Delta\nu_l \frac{1 -
e^{-\eta_l t}}{t} \label{mLinesDef}, 
\end{eqnarray}
with
\begin{equation}
\eta_l \equiv  \frac{\chi_l}{\sigma_T n_e} \hspace{1in}
t \equiv \frac{\sigma_T n_e v_{\rm th}}{dv_R/dR}, \label{tDefinition}
\end{equation}
where $\nu$ is the photon frequency, $F_l$ is the local (transmitted)
flux in the line at the frequency of line number $l$, $v_{\rm th}$ is
the sound speed in the gas, $\Delta \nu_l = \nu v_{\rm th}/c$ is the
thermal line width, and $\eta_l$ compares the opacity of the line (for
a given ionization state of the gas) to the electron opacity,
representing all of the atomic physics in the radiative acceleration
calculation.  The last remaining variable, $t$, is often called the
``effective electron optical depth'' and encodes the dynamical
information of the wind in the radiative acceleration calculation.
This dynamical information is important because in an accelerating
medium, one must also account for the Doppler shift of the atomic line
absorption energy in the accelerating gas relative to the emitted line
photon's energy: beyond the Sobolev length, $v_{\rm th}/(dv_R/dR)$,
included in $t$, a line photon will be Doppler-shifted out of the
thermal width of the absorption line and can escape the gas
\citep*[see][]{Sobolev58, CAK75, MW99}.

The above-mentioned force multipliers are calculated using the
resonance line data of \citet{Verner96} with solar abundances.  Since
$M_{\rm cont}$, the continuum multiplier, depends only on the
ionization state, it is tabulated solely as a function of height in
the wind.  In contrast to $M_{\rm cont}$, the line multiplier ($M_{\rm
line}$) is tabulated for a range of values of the parameter $t$.
Later, when calculating the equation of motion for the wind, the local
velocity gradient is used to compute the actual value of $t$, which is
then used to linearly interpolate the table of $M_{\rm line}$ and then
evaluate the radiative acceleration.

\begin{figure*}[ht]
\includegraphics[width=13.5cm,angle=-90]{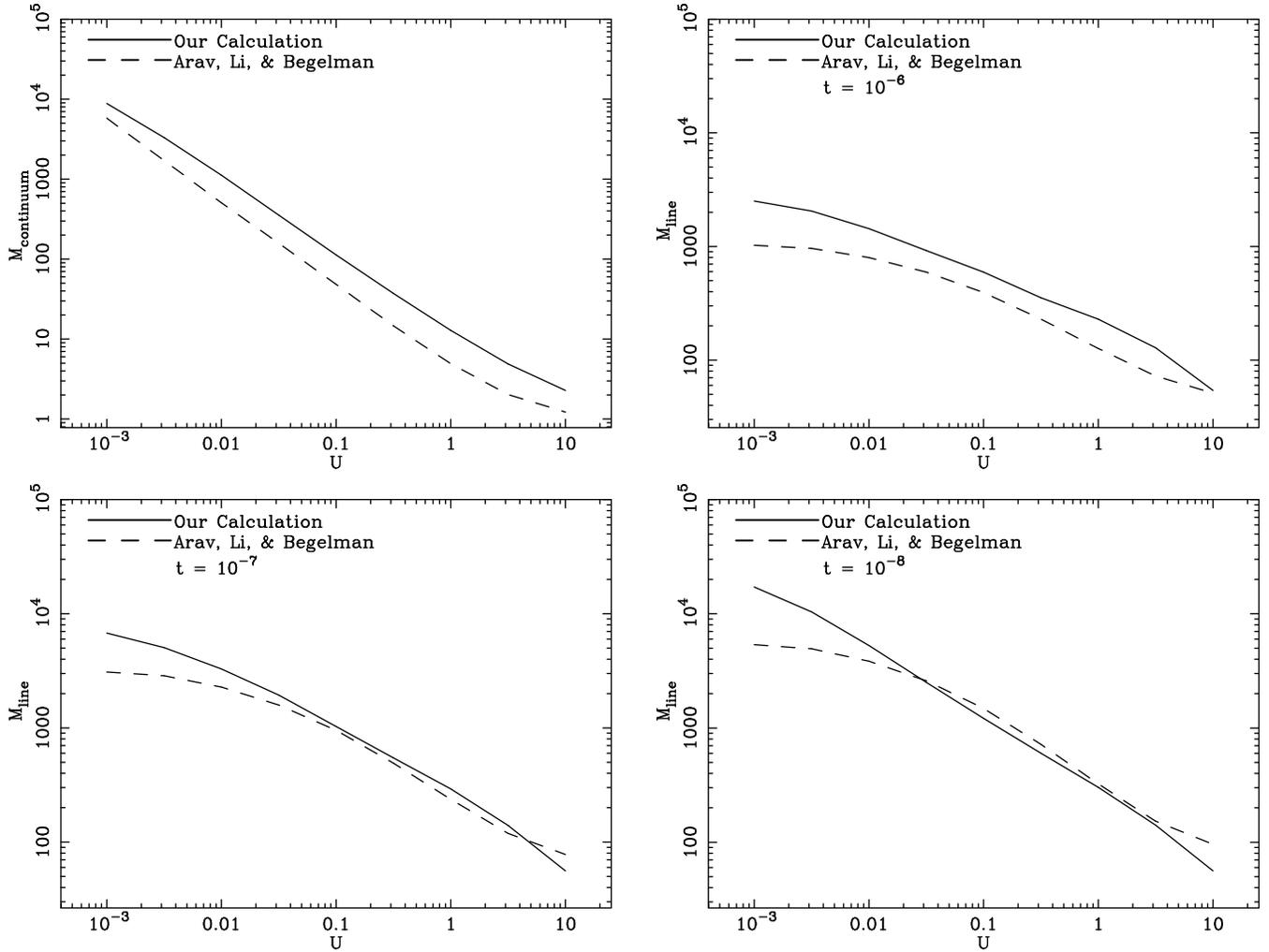} 
\caption{Comparison between the force multiplier calculation in this
code ({\em solid lines}) and the fits of \citet[{\em dashed
lines}]{Arav94}.  The plot in the upper left is the comparison of the
continuum force multiplier, as a function of the ionization parameter,
$U$.  The next three panels show a comparison of the line force
multiplier (which is also a function of $t$, besides $U$) as a
function of $U$ for three different values of $t$, given at the top of
each panel.
\label{forceMultCompare}}
\end{figure*}

The force multiplier computation has been tested against
\citet{Arav94}, who also calculated radiative acceleration from
photoionization simulations.  Figure~\ref{forceMultCompare} compares
these new calculations results against their fits (noting that there
is a typo in their eq.~[2.9]; Z.-Y. Li, personal communication), where
the force multipliers as a function of the ionization parameter $U$ is
presented ($U$ is the ratio of hydrogen-ionizing photon density to
hydrogen number density $n$, given by $U \equiv Q/4{\pi}nR^2c$, where
$Q$ is the number of incident hydrogen-ionizing photons per second,
and $R$ is the distance from the continuum source).  Overall, good
agreement is found, especially considering that \citet{Arav94} point
out that their fit deviates from their calculations at low values of
$U$.  The increase in the newly-calculated continuum force multiplier
over \citet{Arav94} is most likely due to the different continuum
opacity database included in Cloudy 96 compared to the code (MAPPINGS)
that was used in \citet{Arav94}.  The multiplier values and trends
with ionization parameter are still clearly very similar, however.

\subsubsection{Non-Sobolev Effects}\label{nonSobolevEffects}

Simply using the Sobolev length, $v_{\rm th}/(dv_R/dR)$, in Equation
\ref{mLinesDef} can be misleading.  In early simulations, we found
that this Sobolev length could, in regions where the gas is slowly
accelerating, be much larger than the physical size of the shield and
wind combined.  This is clearly not physical, so a simple non-Sobolev
method was employed to calculate the size of the combined shield and
wind column.

First, for the wind, the length of the absorbing column is simply
limited to the minimum of the wind's Sobolev length and its true
(physical) length.

Second, for the shield, the length of the absorbing column is given by
the minimum of the shield's Sobolev length\footnote{The Sobolev length
calculation for the shield does take into account the offset in
velocity between the wind and shield, which is important since the
wind is radiatively accelerated in addition to the magnetic
acceleration that the wind and shield share.}, its true physical
length, and the average length of the column that absorbs photons at
the wavelength of the dominant accelerating atomic lines.  This last
length-scale requires more explanation.  To calculate this length, the
top 20 line force multipliers (for individual atomic lines) are found
for each polar angle $\theta$, and for each of those lines, the column
of each particular ion in the shield is read from the Cloudy
simulations.  An average shielding column for all of the high-opacity
lines is then calculated.  To test this approximation, we have used
lists of the top 10, 20, and 50 transitions in the wind and used them
to calculate the limiting column in the shield.  Changing the number
of lines included does not significantly change the final wind
solutions, especially since this effect is only important at the
extreme base of the wind.  Without considering all of these
constraints on the size of the absorbing column, the Sobolev length
can significantly overestimate the optical depth in the shield (even
overestimating the physical length of the columns, for small
accelerations), which results in the line driving acceleration
dropping below the continuum acceleration.  Other non-Sobolev effects
(such as line blanketing within the shielding gas or the wind) are not
considered in this model; for a consideration of these effects, see
\citet{CN03a}.

With a length-scale found from the sum of the wind and shield lengths
found above, that length-scale is then substituted for the Sobolev
length in the calculation of $M_{\rm lines}$ in Eq.~\ref{mLinesDef}.
Including this physical length of the wind and shield will introduce
dependences on the sizes of the wind and shield columns, which will be
examined in \S\ref{modelDependences}.

\subsection{Integrating the Euler Equation for the Wind}\label{windEqMtn}

Given the results from the ionization and radiative acceleration
calculations, the next step is to solve for the effect of radiative
acceleration on the wind, taking the magnetocentrifugal wind model
already computed and augmenting its acceleration with radiative
forces.  To do this, the full equation of motion (the Euler equation)
is integrated along the streamline of the self-similar wind solution,
recording $\Gamma(\theta)$, the radiative acceleration.
$\Gamma(\theta)$ is then input into the self-similar model in the
subsequent iteration (see Appendix A; specifically,
eqn. \ref{gammaInSelfSim}).

In its simplest form, Euler's Equation is given by
\begin{eqnarray}
\rho \left( \frac{\partial \mathbf{v}}{\partial t} + (\mathbf{v} \cdot
\nabla)\mathbf{v} \right) = \sum \mathbf{F}_i, \label{simpEuler}
\end{eqnarray}
where the $\mathbf{F}_i$ represent the various forces included.

As we are interested in steady-state winds, $\frac{\partial
\mathbf{v}}{\partial t}$ in Eq.~\ref{simpEuler} is set to zero.  As
already mentioned, gravitational, radiation, and Lorentz forces are
included, while thermal terms are not.  This yields the expression
below, with the magnetic force term is split into pressure and tension
components.
\begin{eqnarray}
\rho (\mathbf{v} \cdot \nabla)\mathbf{v} & = & - [1 - \Gamma(\theta)]
\frac{G M \rho}{R^2}\hat{R} - \frac{1}{8 \pi} \nabla B^2 + \frac{1}{4
\pi} (\mathbf{B} \cdot \nabla)\mathbf{B}
\end{eqnarray}

For these calculations, it is more intuitive to integrate the equation
of motion along the flow already given by the magnetocentrifugal wind
solution.  Therefore, taking the dot product of Euler's equation with
$\hat{s}$, which is defined as the direction along the flow, and
expanding and simplifying the left-hand side of the equation, one
finds:
\begin{eqnarray}
(\mathbf{v} \cdot \nabla \mathbf{v}) \cdot \hat{s} & = & v_p 
\frac{\partial v_p }{\partial s} -
\frac{v_{\phi}^2}{r} \sin \theta_F,
\end{eqnarray}
where
\begin{eqnarray}
\theta_F & \equiv & \tan^{-1} \left( \frac{dr}{dz} \right).
\end{eqnarray}
In the same way, if $\theta$ is defined via
\begin{eqnarray}
\theta \equiv \tan^{-1} \left( \frac{r}{z} \right),
\end{eqnarray}
the gravitational term can be written as
\begin{eqnarray}
- [1 - \Gamma(\theta)] \frac{G M}{R^2}\hat{R} \cdot \hat{s} & =
  & - [1 - \Gamma(\theta)] \frac{G M}{r^2 + z^2} \cos (\theta - \theta_F).
\end{eqnarray}
Next, we take the dot product of $\hat{s}$ with the magnetic terms to
find
\begin{eqnarray}
\left[ -\frac{1}{8 \pi \rho} \nabla B^2 + \frac{1}{4 \pi \rho}
(\mathbf{B} \cdot \nabla)\mathbf{B} \right] \cdot \hat{s} & = &
-\frac{B_{\phi}}{4 \pi \rho r} \frac{\partial (r B_{\phi})}{\partial s}.
\end{eqnarray}
Combining all of those terms, the full Euler Equation is obtained:
\begin{eqnarray}
v_p \frac{\partial v_p }{\partial s} - \frac{v_{\phi}^2}{r} \sin
\theta_F  & = & - [1 - \Gamma(\theta)] \frac{G M}{(r^2 + z^2)} \cos
(\theta - \theta_F) \nonumber \\ & & - \frac{B_{\phi}}{4 \pi \rho r} \frac{\partial (r
B_{\phi})}{\partial s}.\label{fullEulerEqn}
\end{eqnarray}

This equation is still dependent on $v_{\phi}$, however, which can be
eliminated by appealing to the induction equation and $E_{\phi} = 0$
for such axisymmetric systems \citep[see, e.g.,][]{KP00}, to yield a
relation between $v_p$ and $v_{\phi}$:
\begin{eqnarray}
v_{\phi} & = & \frac{v_p B_{\phi}}{B_p} + \Omega r.
\end{eqnarray}

Substituting this expression into the Euler Equation yields:
\begin{eqnarray}
\lefteqn{v_p \frac{\partial v_p }{\partial s} - \left(\frac{v_p B_{\phi}}{B_p}
+ \Omega r \right)^2 \frac{\sin \theta_F}{r} = } \nonumber \\ & &   - [1 -
\Gamma(\theta)] \frac{G M}{(r^2 + z^2)} \cos (\theta - \theta_F)
-\frac{B_{\phi}}{4 \pi \rho r} \frac{\partial (r B_{\phi})}{\partial s}.
\end{eqnarray}

To evaluate the effective optical depth $t$, an expression is required
for $dv_R/dR$, the spherical radial gradient of the spherical radial
velocity.  Since the code integrates quantities only parallel to the
flow, approximations to the perpendicular velocity gradients must be
used.  Assuming that the derivatives of $\theta$ and $\theta_F$ with
distance along the streamline are small (verified \textit{a
posteriori} to be true):
\begin{eqnarray}
\frac{dv_R}{dR} & = & \hat{R} \cdot \nabla v_R, \\
& = & \hat{R} \cdot \nabla [v_p \cos (\theta - \theta_F)], \\
& \approx &  \cos (\theta - \theta_F)
\left(\frac{dv_p}{dr} \sin \theta + \frac{dv_p}{dz} \cos \theta \right), \\
& \approx &  \cos (\theta - \theta_F) \frac{dv_p}{ds} (\sin \theta_F
\sin \theta + \cos \theta_F \cos \theta), \\
& = & \cos^{2} (\theta - \theta_F) \frac{dv_p}{ds}.
\end{eqnarray}



This integration procedure has been tested with radiative acceleration
turned off, where it reproduces the original self-similar velocity
profile to within one part in $10^{5}$.  With the radiative
acceleration turned on, the entire code has repeatedly converged
within approximately eight iterations to an equilibrium magnetic wind
structure (see Fig.~\ref{windIterations}).  These tests show that
consistent solutions are found and that radiation pressure does indeed
affect the shielded component of the wind.

\subsubsection{Critical Points}

As with any steady-state wind dynamics problem, one must search for
and consistently pass all critical points \citep[e.g.,][]{V00}.
Critical points mark the location in the wind where the flow speed is
equal to the speed of information propagation in the wind, and mark
locations in the solutions where solutions branches, or roots, meet;
in the case of radiatively accelerated winds, the location of the
critical point is set by the information propagation speed of
radiative-acoustic wave, or Abbott speed \citep[see,
e.g.,][]{Abbott80, MW99}.  In integrating the equation of motion for
the radiatively-accelerated wind, this code searches for critical
points by looking for multiple roots in the solution to the equation
of motion.  However, no critical points due to radiative acceleration
are present in any of the wind solutions we have found (the
magnetocentrifugal wind does, however, always pass through its own
Alfv\'en critical point).

To check this result, we have duplicated the work of \citet[hereafter,
FS99]{FS99}, verifying that for simple wind geometries and without
magnetocentrifugal acceleration, the integration code does indeed
encounter a radiative critical point as predicted and found by FS99.
In particular, for the field geometries and forces used in FS99, we
have found identical solutions to both their analytical and numerical
calculations.

We then gradually add, to the FS99 model, new components that are
present in our new calculations.  When the centrifugal acceleration
and the enforced corotation near the base of the wind are introduced
into the framework of FS99, the centrifugal acceleration overwhelms
the acceleration of the second root that was present in FS99.
Therefore, only a single root is found, and with only a single root, a
critical point cannot be present in our solutions.  Another way to
check this is to examine the limit of winds launched at very large
angles to the accretion disk.  Indeed, for those large angles, the
radiative acceleration begins to dominate the centrifugal
acceleration, and the critical point reappears.  Therefore, for the
geometry of these magnetocentrifugal winds, where the angle of the
outflow to the accretion disk surface is less than 60$^\circ$, no
radiative critical point is expected within these solutions: a
radiative critical point will not be present when centrifugal
acceleration is dominant.

\subsection{Model Assumptions and Limitations}\label{assumptions}

This model includes simplifying assumptions about the outflow in order
to make these calculations possible.  In this subsection, the
assumptions and limitations of the model are summarized, as are the
reasons for allowing those assumptions.

First and foremost among these assumptions is self-similarity. While
enabling a relatively quick and flexible model that can be used to
survey a wide variety of outflows, this assumption does impose
constraints on the dynamics of the model.  However, the assumption of
self-similarity is essential to the magnetocentrifugal wind solution,
as it simplifies the complicated MHD equations.  Simultaneously, the
radial self-similarity accommodates, very easily, the radial geometry
for the photoionization simulations. 

The above-mentioned Cloudy photoionization simulations assume a static
medium, which is an approximation as well.  As we show in
\S\ref{photoSims}, however, this approximation is valid for the these
wind models.


Since the accretion disk is a boundary condition in these models,
these winds are assumed to be loaded with matter from the disk.
Accurate models of the accretion disk structure are beyond the scope
of this paper, so it is assumed that the full mass outflow rate of the
the wind is indeed input onto the magnetic fieldlines at the accretion
disk surface.  In addition, the matter that is loaded onto those
fieldlines is assumed to flow supersonically, i.e., the gas has
already passed the sonic point in the flow.  This is done to simplify
the magnetocentrifugal wind equations and retain the basic model as
outlined in BP82; as such, the same asymptotic expansions near the
disk (from BP82) are utilized here.  This treatment of the wind has
been checked in several different ways.  First, the magnetic pressure
in the wind is indeed greater than the thermal pressure throughout the
entire wind.  Also, the wind's final velocities are much greater than
the sound speed at the base of the wind, showing that thermal effects
are negligible in determining the final wind velocities.  Finally, the
lowest speeds found in the wind model are of order 20\% of the sound
speed, and such low Mach numbers are found only very near the disk, at
the base of the wind.  Given the above evidence of the dominance of
magnetic fields and radiative acceleration, the ``cold-wind''
approximation is valid for these calculations.

Also, in the calculation of the radiative acceleration of the wind, an
approximation to the velocity gradient along spherical rays is
required.  This approximation is necessary because the Euler
integration for the wind yields velocity gradients only along the flow
(the poloidal velocity gradients), and thus the other components must
be approximated geometrically (see \S\ref{windEqMtn}).


	The approximations and limitations outlined above do constrain
the use of this wind model, but in making these compromises, a
versatile tool can be developed to study the chosen geometries and
forces.

\section{Radiation Transfer Within a Fiducial Magnetocentrifugal Wind}\label{fiducialModel}

For definitiveness, this model is first employed to examine the
radiative transfer within one fiducial irradiated magnetocentrifugal
wind.  For the purpose of this paper, `fiducial' is defined to
indicate the parameters listed in Table~\ref{paramTable}.  These
parameters are not meant to represent a proposed model for any one
particular AGN, but to define a starting point from which to examine
the structure of these outflows as well as the dependences of the
outflows on the model parameters.  For instance, the shielding column
of $N_{\rm H, shield,0} = 10^{23}$~cm$^{-2}$ (where $N_{\rm H,
shield,0}$ represents the value of $N_{\rm H, shield}$ at the base of
the wind, i.e. just above the accretion disk surface) is chosen
because it displays an amount of radiative acceleration between the
extremes of the smaller and larger columns that will be tested.
Similarly, the radiative wind column that is defined is again between
the extremes of nearly optically thick $N_{\rm H,rad,0} =
10^{23}$~cm$^{-2}$ and very optically thin $N_{\rm H,rad,0} =
10^{19}$~cm$^{-2}$.  Studying such a model first will help bring into
focus important issues concerning the interplay of dynamics and
photoionization, and represents a foundation from which one can
explore the parameter dependencies of the model.

\begin{figure}[ht]
\begin{center}
\includegraphics[width=8cm]{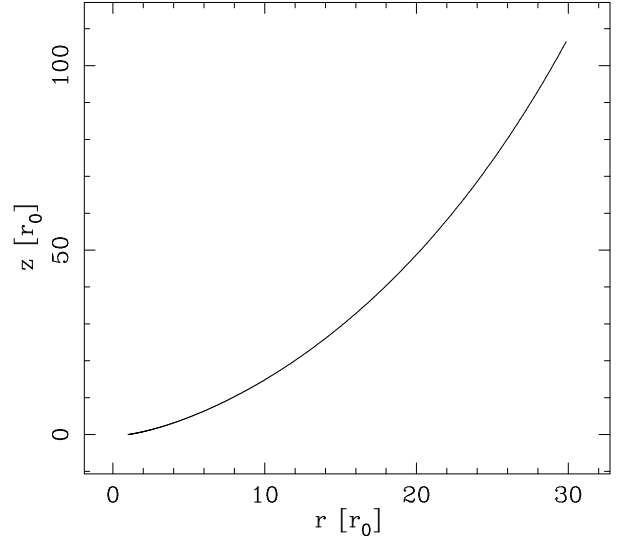}
\caption{Poloidal wind streamlines in the fiducial model.  Both the
  cylindrical radial coordinate, r, and the height, z, are given in
  terms of the launch radius, $r_0$.  Note that the wind is still
  somewhat collimated, since the radiative acceleration input is
  fairly low for $L/L_{\rm Edd} = 0.01$.  
\label{windStreamline}}
\end{center}
\end{figure}

\begin{figure}[ht]
\begin{center}
\includegraphics[width=8cm]{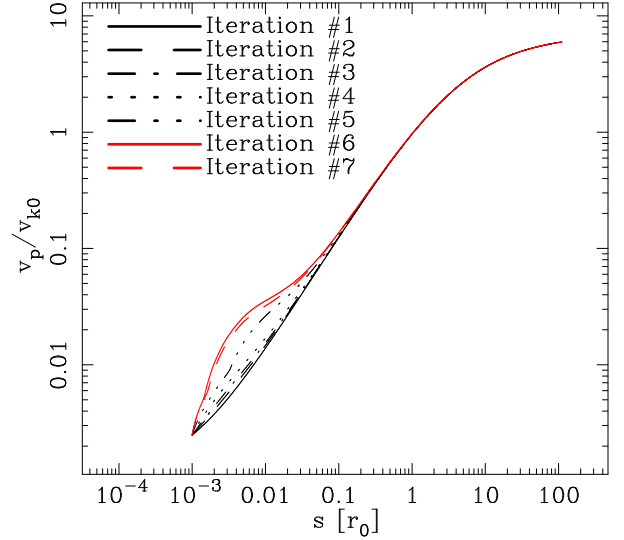}
\caption{Poloidal velocity for the fiducial wind as a function of
  distance along the flowlines, showing the evolution of the velocity
  over several iterations of the code.  Line driving does
  significantly affect the magnetocentrifugal wind near the base, but
  magnetocentrifugal driving dominates at larger distances for this
  Eddington ratio.  (In the first iteration in this plot, only
  magnetocentrifugal driving has been considered, so the ``Iteration
  \#1'' curve shows the pure magnetocentrifugal wind case.)
  \label{windIterations}}
\end{center}
\end{figure}

This fiducial model was therefore run with the parameters given in
Table~\ref{paramTable}, and after eight iterations, converged to the
final wind structure.  The results for the fiducial model are shown
in Figures~\ref{windStreamline} though \ref{windIterations}.  In these
figures, an overview of the equilibrium state of this model is
presented.  The results displayed in these plots are discussed in
detail below.

\begin{figure}[ht]
\begin{center}
\includegraphics[width=8cm]{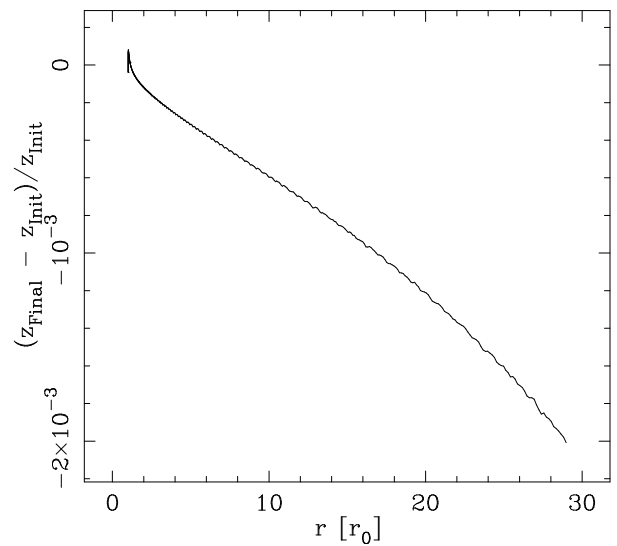}
\caption{Fractional difference in streamline height as a function of
  radius between the final iteration and first iteration of the
  fiducial model.  The small different in height emphasizes that the
  structure of the magnetocentrifugal wind is not significantly
  modified for the fiducial model's low $L/L_{\rm Edd} = 0.01$.
\label{windStreamlineDifference}}
\end{center}
\end{figure}

First, Figure~\ref{windStreamline} shows the height of the poloidal
streamline as a function of radius in units of the launching radius.
In addition, to illustrate the small difference in geometry between
the final and initial wind models,
Figure~\ref{windStreamlineDifference} gives the fractional change in
height as a function of radius.  Both of these figures show that the
wind still maintains a collimated state, achieving a height of $z/r_0
\sim 100$ (where $r_0$ is the launching radius) at a cylindrical
radius of only $r/r_0 \sim 30$.  So, in this fiducial model, despite
the input from radiative acceleration, the wind maintains this
streamline with only small changes in the structure of the wind
throughout all iterations (see Fig.~\ref{windStreamlineDifference}).
Thus, for the case of the fiducial model with $L/L_{\rm Edd} = 0.01$,
this added acceleration does not significantly affect the structure of
the magnetocentrifugal outflow.  These models do show changes in the
velocity structure of the wind near the disk surface (as will be shown
in Fig.~\ref{windIterations}), but the poloidal wind structure does
not change significantly: on the scale of Figure~\ref{windStreamline},
the streamlines of the initial, purely magnetocentrifugal streamline
would lie on top of the streamline shown.

\begin{deluxetable*}{lll}
\tablewidth{0pt}
\tableheadfrac{0}
\tablecaption{Parameters adopted for the `fiducial' model in this
study.
\label{paramTable}} 

\tablehead{
\colhead{Parameter} & \colhead{Fiducial Value} & \colhead{Parameter
Description} } 

\startdata 
$n_0$ & 
      $10^9~{\rm cm}^{-3}$ & 
      initial density of the wind at the launch radius \\ 
$N_{\rm H,shield,0}$ & 
      $10^{23}~{\rm cm}^{-2}$ & 
      gas shielding column at the base of the wind \\
$N_{\rm H,rad,0}$ & 
      $10^{21}~{\rm cm}^{-2}$ & 
      gas column, behind the shield, that is radiatively accelerated \\
$M_{\bullet}$ & 
      $10^8 M_{\sun}$ & 
      mass of the central black hole\\
Incident Spectrum & 
      \citet{RE04} & 
      Spectrum for the central continuum \\
$L_{\rm continuum}$ & 
      $0.01~L_{\rm Edd}$ & 
      luminosity of the central continuum\\
$r_0$ & 
      $3 \times 10^{16}~{\rm cm}$ & 
      launch radius of the wind \\
$\kappa$ & 
      0.03 & 
      dimensionless ratio of mass flux to magnetic flux in the wind\\
$\lambda$ & 
      30.0 & 
      normalized total specific angular momentum of the wind\\
$b$ & 
      1.5 & 
      power-law describing variation of density with spherical radius in the wind: $n \propto R^{-b}$\\
Dust in Wind & 
      No & 
      presence of dust in the wind \\
\enddata
\end{deluxetable*}

\begin{figure*}[t]
\begin{center}
\includegraphics[width=11cm,angle=-90]{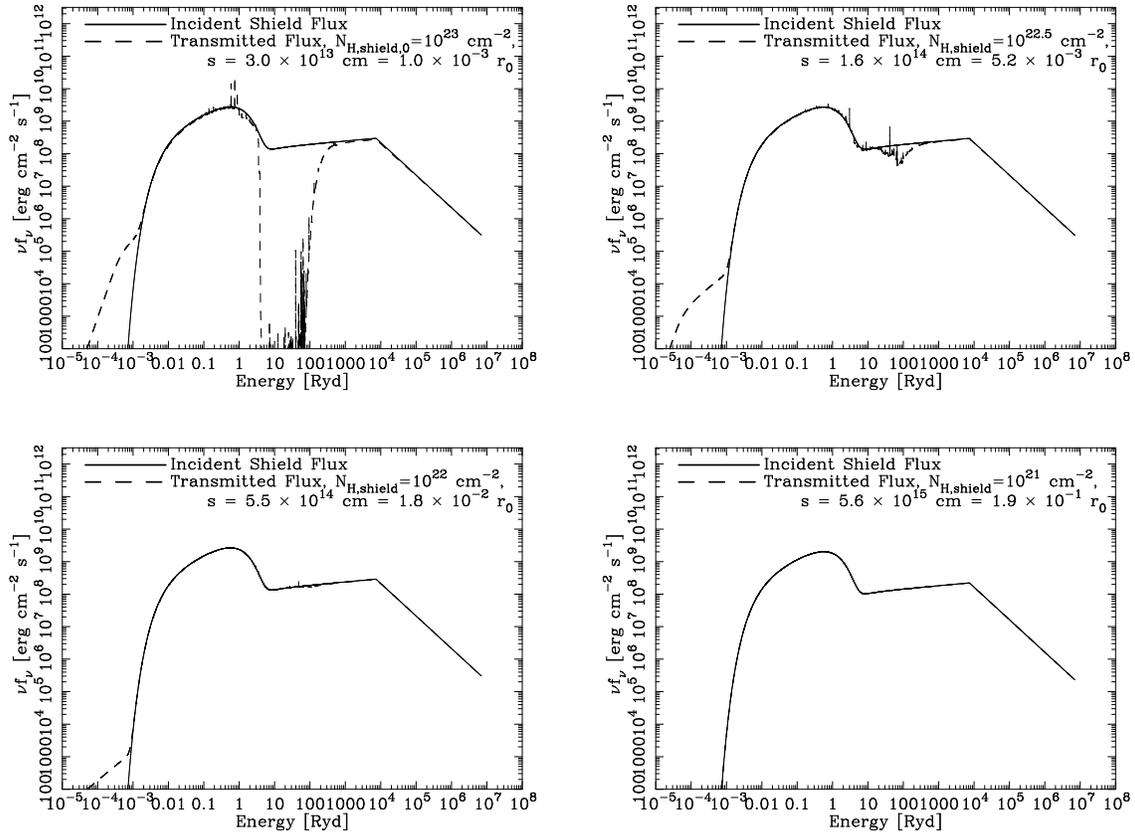}
\caption{Both the incident continuum on the shield and the transmitted
  continuum through the shield is displayed.  Note the rapid decline
  in the shield's column density as a function of polar angle: $N_{\rm
  H, shield,0} = 10^{23}$~cm$^{-2}$ lies at the polar angle $\theta =
  89.9^\circ$, $N_{\rm H, shield} = 10^{22.5}$~cm$^{-2}$ at $\theta =
  89.8^\circ$, $N_{\rm H, shield} = 10^{22}$~cm$^{-2}$ at $\theta =
  89.4^\circ$; the radiation encounters $N_{\rm H, shield}$
  $10^{21}$~cm$^{-2}$ at only $\theta = 85.1^\circ$.  Also indicated
  for each column is the distance along the flowline ($s$) where that
  column occurs, given both in cm and in units of the launching
  radius, $r_0$. 
\label{transmittedCont}}
\end{center}
\end{figure*}

At logarithmically-spaced co-latitudinal angles along the streamline,
Cloudy photoionization simulations are run to determine the
photoionization state of the gas, as well as the radiative transfer
through the shield and wind.  Changes in the continuum transmitted
through the shield are shown in Figure~\ref{transmittedCont}; the
various plots show the simulated continuum at various heights in the
shield corresponding to the indicated columns.  As the shielding
column decreases as a function of height above the disk, the shield
transmits progressively more and more of the ionizing radiation.  This
plot also displays how rapidly the column drops as a function of
height above the disk: $N_{\rm H, shield,0} = 10^{23}$~cm$^{-2}$
occurs at $\theta = 89.9^\circ$, $N_{\rm H, shield} =
10^{22.5}$~cm$^{-2}$ at $\theta = 89.8^\circ$, $N_{\rm H, shield} =
10^{22}$~cm$^{-2}$ at $\theta = 89.4^\circ$, and $N_{\rm H, shield}$
drops to $10^{21}$~cm$^{-2}$ at $\theta = 85.1^\circ$.

\begin{figure*}[ht]
\begin{center}
\includegraphics[width=5.5in]{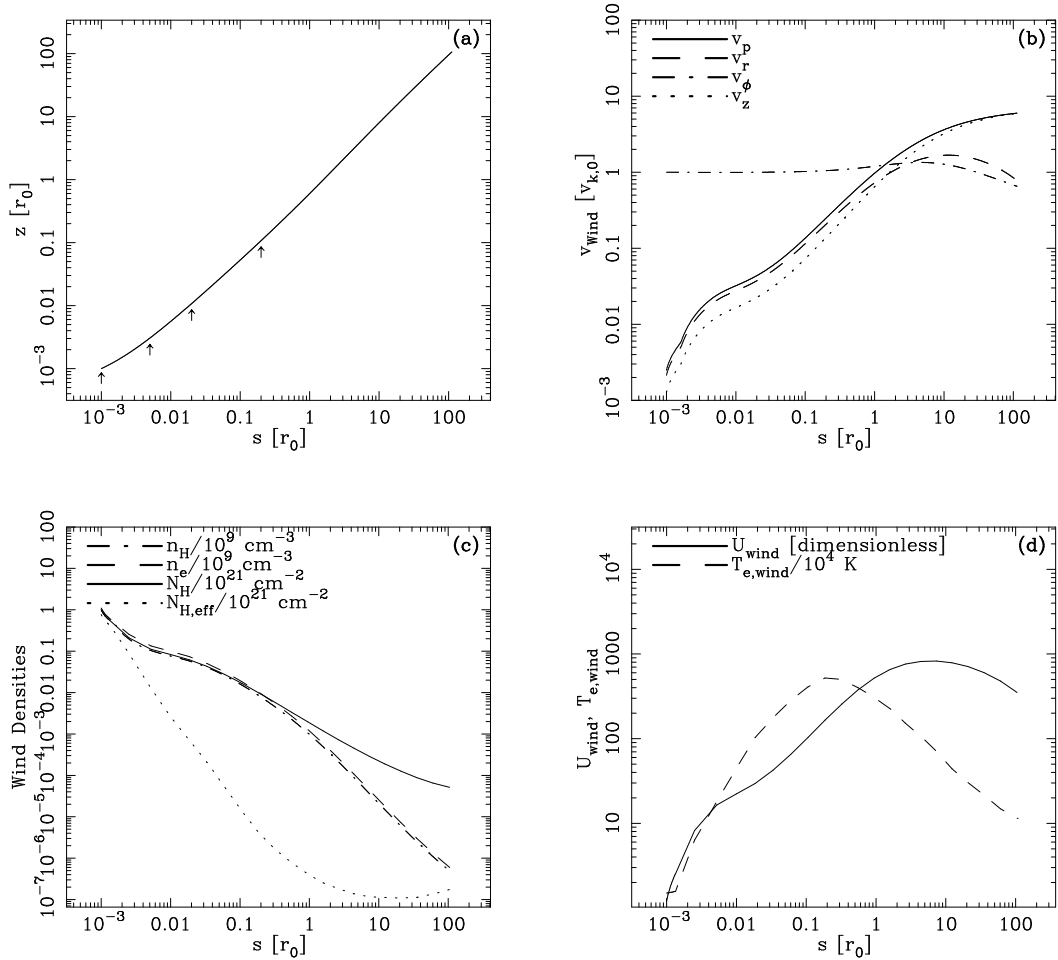}
\caption{The geometry, dynamics, and photoionization state of an
  illuminated, radiatively-accelerated magnetocentrifugal wind.  Pane
  (a) shows the height of the wind (in units of the launching radius)
  with respect to distance along the flowline, $s$.  The arrows in
  pane (a) give the positions of the columns shown in
  Figure~\ref{transmittedCont} ($10^{23}$, $10^{22.5}$, $10^{22}$,
  $10^{21}$~cm$^{-2}$; decreasing with increasing $s$ and $z$).  Pane
  (b) shows the various components of the velocity: poloidal, radial,
  azimuthal, an vertical, respectively, in units of the Keplerian
  speed at the base of the flow (for the fiducial model, that base
  Keplerian speed, $v_{\rm k,0}$, is $6.65 \times 10^3$~km s$^{-1}$).
  Pane (c) displays various densities: the total hydrogen and electron
  number densities as well as the total hydrogen column and the
  effective Hydrogen column (defined as the column of cold, neutral
  gas of solar abundances that would produce the same obscuration at 1
  keV).  All densities are plotted in units of the number density or
  total column density at the base of the wind.  Pane (d) shows both
  the electron temperature in the gas and the ionization parameter, $U
  = n_{\gamma}/n_{H}$ where $n_{\gamma}$ is the number density of
  hydrogen-ionizing photons.
\label{fiducialWind}}
\end{center}
\end{figure*}

The flux transmitted through the shield then illuminates the
radiatively accelerated wind; results from the photoionization
simulations for the wind are presented in Figure~\ref{fiducialWind},
where the streamline, velocity, density, ionization parameter and
temperature in the wind are plotted.  In Figure~\ref{fiducialWind}a,
the height of a wind streamline as a function of distance along the
streamline (labeled $s$, given in units of the initial radius, $r_0$)
is shown; this plot simply recasts the structure of the flowline shown
in Figure~\ref{windStreamline} in terms of $s$ for comparison with the
remaining plots.

In Figure~\ref{fiducialWind}b, velocities along the streamline are
plotted, showing not only the rapid acceleration in the wind, but also
comparing the components of the wind's velocities.  All velocities are
given in units of the Keplerian velocity at the base of the wind,
$v_{\rm k,0}$ ($v_{\rm k,0} = 6.65 \times 10^3$~km s$^{-1}$ for the
parameter values in Table~\ref{paramTable}).  This plot shows that the
vertical velocity is dominant in these winds at large distances (again
showing the wind is somewhat collimated), with the radial and
azimuthal velocities becoming approximately equal far from the
launching radius of the outflow.  Near the very base of the disk, the
radial velocity quickly dominates both the azimuthal and vertical
velocities (qualitatively similar to the velocity structure calculated
for radiatively-dominated flows as in MCGV95).  Most importantly,
though, we note the extraordinarily rapid acceleration of the gas from
the disk; such acceleration is a hallmark of both magnetocentrifugal
as well as radiatively-dominated winds, which usually accelerate to
their terminal velocities in a distance on the order of their
launching radius.

\begin{figure*}[ht]
\begin{center}
\includegraphics[width=5.5in]{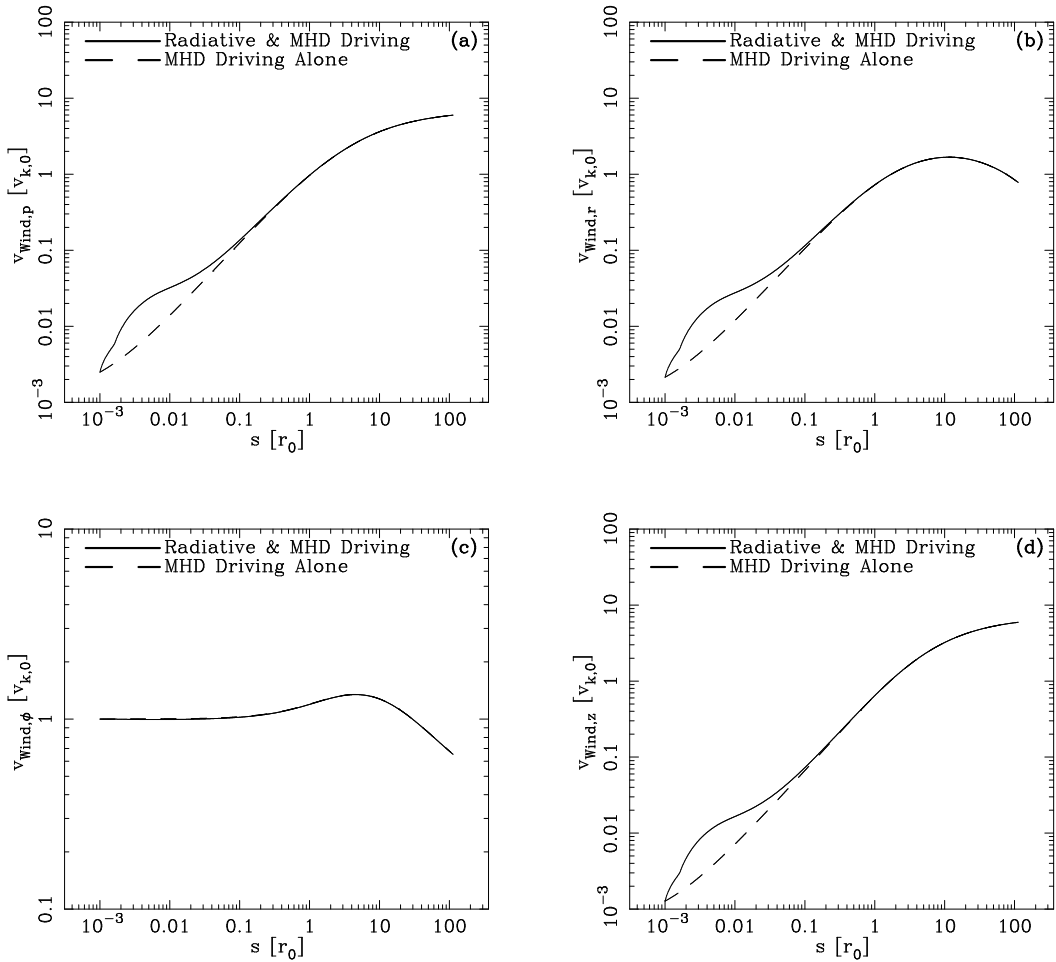}
\caption{Variation in all components of the velocity between the
  initial, pure magnetocentrifugal wind model and the final
  magnetocentrifugal \& radiatively-driven model.  The poloidal
  velocity (shown in pane a) shows the effects of radiative
  acceleration (increasing the velocities near the base of the wind by
  approximately a factor of two for this low $L/L_{\rm Edd} = 0.01$),
  as do the poloidal velocity's components, the radial and vertical
  velocities (panes b and d).  The azimuthal component of the velocity
  is relatively unaffected (pane c).
\label{fiducialWindVelDiff}}
\end{center}
\end{figure*}

\begin{figure}[h]
\begin{center}
\includegraphics[width=8cm]{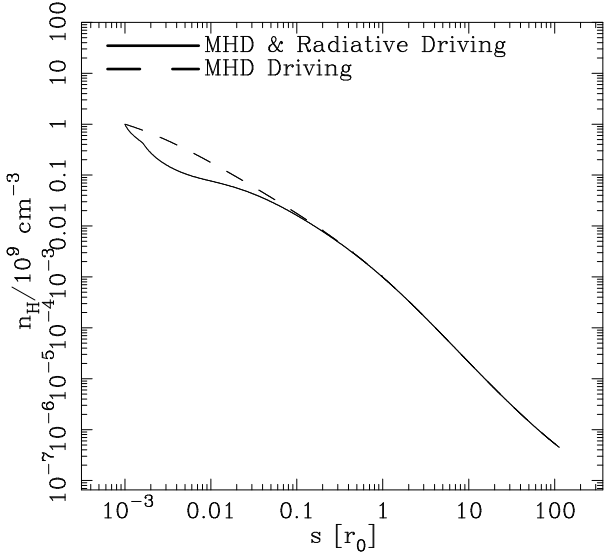}
\caption{Variation in density between the initial, pure
  magnetocentrifugal wind model and the final magnetocentrifugal \&
  radiatively-driven model.  Shown here is the change in the hydrogen
  density with height, but the electron density and columns change
  similarly: the added radiative acceleration yields a drop in density
  near the base of the disk, as can be seen in all of the various
  density measurements.
\label{fiducialWindDensityDiff}}
\end{center}
\end{figure}

Due to mass conservation, the extremely rapid acceleration of the wind
causes a sharp drop in both the number density and the column density
with height above the disk: both the number density and column density
immediately drop by three orders of magnitude as the wind rises above
the disk.  This is displayed in Figure~\ref{fiducialWind}c. This
overall drop in density is extremely important for the ionization
state of the wind, not only for the observational ramifications (i.e.,
what ions are present in various parts of the wind) but for the
acceleration of the wind as well, as will be shown very shortly.
Figures~\ref{fiducialWindVelDiff} and \ref{fiducialWindDensityDiff}
show in more detail how the radiative acceleration leads to a
substantial change (by approximately a factor of two) in both velocity
and density with height near the disk surface.  The changes in
velocity for the pure magnetocentrifugal wind as compared to the
magnetocentrifugal and radiatively-accelerated wind is shown in
Figure~\ref{fiducialWindVelDiff}.  The difference in the density
profile for the pure magnetocentrifugal wind as compared to the
magnetocentrifugal and radiatively-accelerated wind is shown in
Figure~\ref{fiducialWindDensityDiff}.  (Note that all densities are
normalized to their value at the base of the wind.)

The corresponding ionization state of the wind and temperature are
shown in Figure~\ref{fiducialWind}d.  Most striking is the dramatic
rise in the ionization parameter as the wind rises above the disk,
which is simply due to the drop in density and in shielding already
mentioned.  The ionization parameter is of prime interest, as the
radiative acceleration in resonant lines is dependent on the number of
atomic lines in the gas; the rapid ionization of the gas prompts
questions about how efficient line-driving will be within this
magnetocentrifugal wind.  In addition, can a wind with such a dramatic
drop in column density form an effective ``shield''?  And for these
models, how does the radiative acceleration then compare to magnetic
acceleration?

\begin{figure}[h]
\begin{center}
\includegraphics[width=6cm,angle=-90]{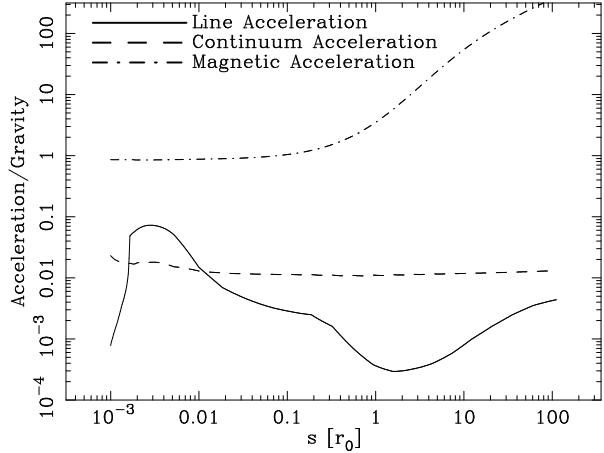}
\caption{Line radiative acceleration, continuum radiative
  acceleration, and magnetocentrifugal acceleration along the gas
  streamlines, compared in the fiducial model.  Note that at the high
  densities and columns near the base of the wind, line-driving
  dominates continuum driving, but that both are less than the
  magnetic driving for a system at $L/L_{\rm Edd} = 0.01$.  At larger
  distances, continuum driving dominates line driving in the more
  highly ionized gas.  As expected, continuum-driving is approximately
  of order the Eddington ratio for the highly-ionized portion of the
  outflow.
\label{lineOverCont}}
\end{center}
\end{figure}

These questions are addressed in Figure~\ref{lineOverCont}.  From the
Cloudy simulations summarized in Figure~\ref{fiducialWind}, both the
bound-free and bound-bound radiative acceleration are calculated; the
resultant acceleration (compared to the local gravity) for the
fiducial model is shown in Figure~\ref{lineOverCont}.  This model
shows that both line-driving and continuum-driving have important
roles to play, with the line-driving dominating the continuum driving
in the high-density, low-ionization part of the wind, and continuum
driving dominating line-driving at larger distances, when the density
is much lower.  It is important to note that for the parameters of the
fiducial model with $L/L_{\rm Edd} = 0.01$, magnetic acceleration is
still much greater than either line or continuum acceleration.
Line-driving is greater than continuum-driving in only part of the
outflow (although this can change with the density at the base of the
wind, as will be shown in \S\ref{nInitVariations}), as the ionization
parameter is low enough only at the base of the wind for significant
numbers of atomic lines to exist.  In addition, radiative driving is
not immediately important at the extreme base of the outflow, because
of the low fluxes that penetrate the columns there.  Meanwhile, the
acceleration due to continuum-driving stays close to the Eddington
ratio, at $\sim0.01$.  This value is reasonable, as most of the
continuum acceleration comes from electron scattering, so that $\Gamma
\sim 0.01$ would be expected.  Minor increases above that value very
near the disk surface are due to bound-free transitions in the portion
of the wind closer to the disk, where $\Gamma_{\rm Continuum}$ rises
to $\sim 0.02$.

As has already been shown, at this low Eddington ratio the structure
of the wind does not change significantly.  However, the velocity at
the base of the outflow is affected.  The change in velocity due to
radiative acceleration is shown in Figure~\ref{windIterations}.  This
figure shows both the poloidal velocity as a function of distance
along the flowline, and the variation in that velocity with iterations
of the model, therefore showing the convergence in the model.
Figure~\ref{windIterations} shows that line-driving near the disk does
significantly accelerate the wind, but magnetic driving determines the
terminal velocity at larger radii.  This figure also displays how the
code converges; the relatively slow convergence during the first four
iterations is the result of the program slowly increasing the
radiative acceleration to the computed value (increased slowly in
order to avoid severely overestimating the radiative acceleration in
the lines and causing sudden deceleration in later iterations).  In
the later iterations, the calculation converges to a final velocity
profile using the full radiative acceleration.  This profile shows the
affect of line-driving near the base of the wind, where the velocity
increases above that of the initial magnetocentrifugal wind.  However,
as already mentioned, the magnetocentrifugal wind (in this model,
where $L/L_{\rm Edd} = 0.01$) still determines the velocity at large
distances.  At those distances, the gas is too ionized to be
appreciably accelerated by line driving.

\begin{figure}[ht]
\begin{center}
\includegraphics[width=6.5cm,angle=-90]{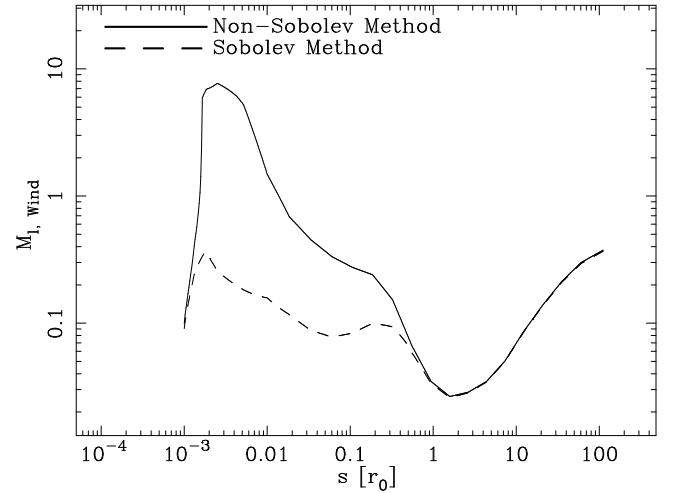}
\caption{Comparison of the line force multiplier, $M_{\rm Lines}$ (see
Eq.~\ref{mLinesDef}), for the case of our non-Sobolev method (defined
in Section~\ref{nonSobolevEffects}) and the strict Sobolev
calculation.  The over-estimate of the absorbing column in the Sobolev
approximation leads to higher opacity in the gas (higher $t$) and thus
lower acceleration.
  \label{figSobolevComp}}
\end{center}
\end{figure}

It is important to note that considering non-Sobolev effects leads to
large changes in $M_{\rm lines}$, the line force multiplier, found in
the above calculations: ``capping'' the Sobolev absorption
length-scale by the actual absorbing column length leads to smaller
optical depths and larger line accelerations.  This is illustrated in
Figure~\ref{figSobolevComp}, where the fiducial model has been
calculated with the Sobolev approximation as well as our non-Sobolev
treatment (see Section~\ref{nonSobolevEffects}).  The difference in
force multiplier is due to the strict Sobolev treatment overestimating
the column for the low-acceleration gas near the base of the wind.
This relatively straightforward modification is very important to
correctly estimate the optical depth, as can be seen in
Figure~\ref{figSobolevComp}.

Overall, in this section, the fiducial model has shown how the wind
velocity, number density, column density, and radiative acceleration
all interact to determine the final state of a magnetocentrifugal
wind.  These components have not previously been self-consistently
combined in a magnetocentrifugal model, and so yield a new look at the
state of these winds.  In addition, the importance of the number
densities and column densities to the final result are most apparent,
and clearly merit further investigation, which will be addressed in
\S\ref{modelDependences}.

\section{Dependence of Wind Structure on Model Parameters}\label{modelDependences}

Having analyzed the fiducial model in detail, and observed how that
model's properties change with height, how sensitive are the trends in
\S\ref{fiducialModel} to those fiducial parameters?  This is an
important question, and one of the key attributes of this self-similar
model is that it allows some flexibility in the selection of initial
parameters.  In this section, we test for variations in the wind by
examining how the wind changes as parameters are modified.

\subsection{Variations with Shielding Column}\label{shieldColumnVariations}

One of the most difficult issues for radiative driving in AGNs is
over-ionization of the wind.  As shown in \S\ref{fiducialModel}, as
the wind accelerates and its density decreases, the magnetocentrifugal
outflow can easily become too ionized to be efficiently accelerated to
escape velocity solely by atomic lines.  This is the problem of the
``shielding gas'' that was mentioned in \S\ref{Intro}: for pure
radiative line-driving, some shielding gas is required to intercept
the X-ray ionizing radiation so that the remaining UV resonant line
photons can be absorbed by the wind and radiatively accelerate it to
the escape velocity.

Some important papers have already been dedicated to examining the
concept of shielding gas, such as \citet[][considering very detailed
photoionization simulations of gas shields with constant column
density]{CN03b} and \citet[][where multidimensional hydrodynamics
simulations with approximate radiative effects are considered]{PK04}.
In contrast, the models presented here include a shield where the
column density varies with height in a shielding wind, and where
detailed photoionization simulations can be employed.

The MHD wind model presented here already launches a wind
magnetocentrifugally, so it is immune to concerns of overionization.
Is it therefore possible for a magnetocentrifugally-driven wind, with
its commensurate drop in column density with height above the disk, to
act as a shield, allowing for more efficient radiative acceleration
beyond it?  We can test this question by simply varying the shielding
column in the fiducial model and checking the radiative acceleration
seen by a wind launched behind the shield.

\begin{figure}[h]
\begin{center}
\includegraphics[width=8cm]{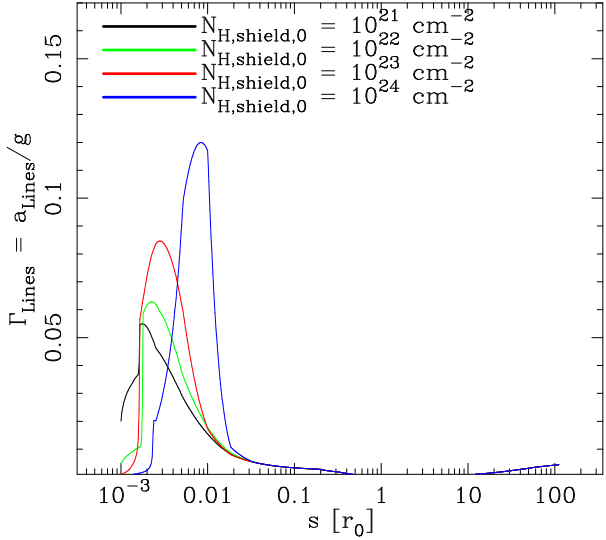}
\caption{Variations in line-driving as the
  magnetocentrifugally-launched shielding column is increased.  With
  greater shielding columns, the wind is less highly ionized, leading
  to higher radiative acceleration near the base of the wind.  Also,
  with increased shielding comes lower flux levels at the base of the
  wind, which displaces the onset of line-driving to larger distances
  from the launch point. (Continuum-driving is approximately constant
  at $a/g \sim 0.01$.)
\label{Gamma_columnScan}}
\end{center}
\end{figure}

As shown in Figure~\ref{Gamma_columnScan}, as the shielding column is
increased from $N_{\rm H,shield,0} \sim 10^{21}$ to $10^{24}~{\rm
cm}^{-2}$, line-driving in the wind increases from $\Gamma_{\rm lines}
\sim 0.05$ to $\Gamma_{\rm lines} \sim 0.12$.  (Recall that for this
model, continuum driving is $\Gamma_{\rm continuum} \sim 0.01$.)  This
shows that, for large shielding columns ($N_{\rm H,shield,0} \sim
10^{24}~{\rm cm}^{-2}$), line driving can be up to an order of
magnitude more effective than continuum-driving at the base of the
wind.  This increase in acceleration is due to the absorption of the
ionizing radiation by the shield, which allows a lower ionization
state in the wind, and more line-driving due to more atomic lines.  It
is also apparent that, as the shielding is increased, the resultant
lower total flux at the base of the wind means that the onset of
significant radiative acceleration is delayed: this accounts for the
offset of maximum radiative acceleration from the disk surface as the
shielding column is increased.

\begin{figure}[h]
\begin{center}
\includegraphics[width=8cm]{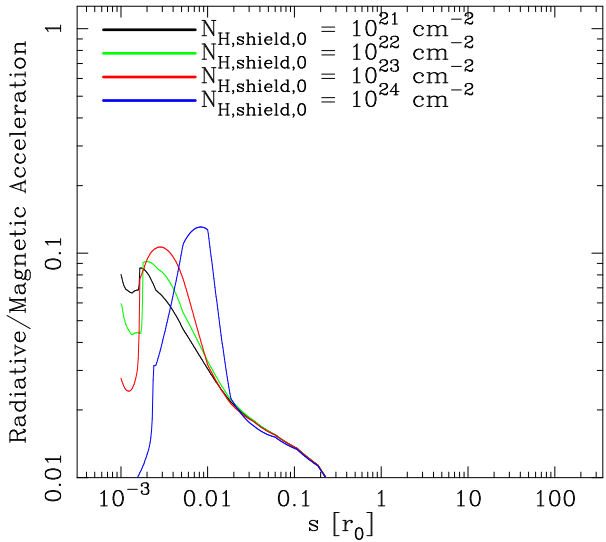}
\caption{As in Figure~\ref{Gamma_columnScan}, but comparing the
  strength of the total radiative acceleration (both line and
  continuum acceleration) to the acceleration due to magnetic fields,
  where both are normalized relative to gravity.  To compare with the
  magnetic field acceleration, the radiative acceleration is scaled by
  a geometrical factor (see Eq.~\ref{fullEulerEqn}) that yields the
  radiative acceleration along the streamline.
\label{Gamma_columnScan_mag}}
\end{center}
\end{figure}

Further, in Figure~\ref{Gamma_columnScan_mag}, the ratio of radiative
acceleration to magnetic acceleration along a streamline is shown.
Since the MHD effects are still supplying most of the acceleration
(with an acceleration roughly equal to and opposite that of gravity),
the ratio of radiative to magnetic acceleration looks much like that
in Figure~\ref{Gamma_columnScan}.  Magnetic effects dominate in these
wind models, even when large columns of shielding are included.

We have therefore shown that a magnetocentrifugal outflow can act as a
shield and increase the efficiency of line-driving in the wind.
However, it can also be seen that line-driving is important in these
models only at the base of the wind.  This arises not only from the
drop in the shield's column density with height above the disk, but
the drop in the wind's density as well (and the commensurate rise of
the ionization parameter).

\subsection{Variations with Initial Density}\label{nInitVariations}

\begin{figure}[h]
\begin{center}
\includegraphics[width=8cm]{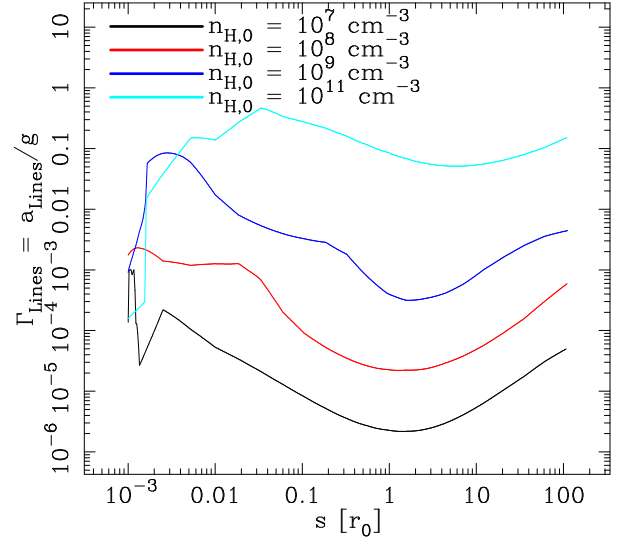}
\caption{Variations in line-driving with changes in the density at the
  base of the wind.  Higher densities lead to smaller ionization
  parameters and much larger acceleration.  (Continuum-driving is
  approximately constant at $a/g \sim 0.01$.)
\label{Gamma_densityScan}}
\end{center}
\end{figure}

\begin{figure}[h]
\begin{center}
\includegraphics[width=8cm]{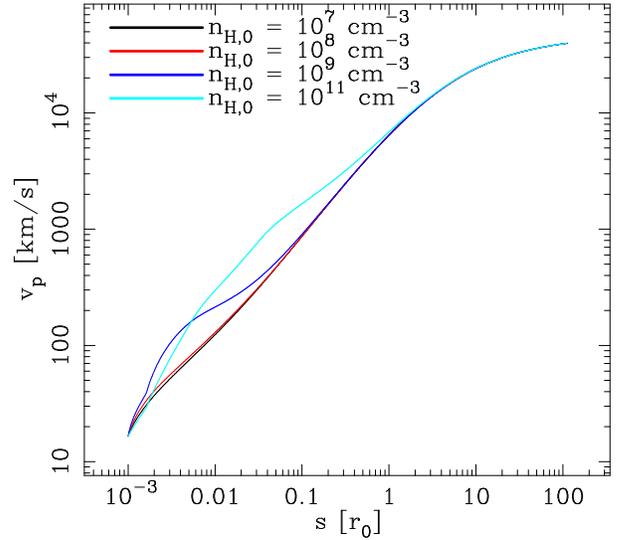}
\caption{As in Fig.~\ref{Gamma_densityScan}, except that the change in
  velocity along the streamline is shown for the variety of densities.
  For this Eddington ratio, the magnetocentrifugal wind still
  dominates, but as the density increases, the importance radiative
  driving increases.
\label{vpVsDensity}}
\end{center}
\end{figure}

Owing to the increase of line-driving with decreasing ionization
parameter, higher accelerations would also be expected at higher
densities.  Thus, we investigate the effect of changes in the initial
density in the wind in Figures~\ref{Gamma_densityScan} and
\ref{vpVsDensity}.

Displaying the effect of a range of initial densities,
Figure~\ref{Gamma_densityScan} shows that line-driving is only
effective in these magnetocentrifugal winds at relatively high
densities.  Since continuum driving is approximately constant (and
relatively independent of density) at $a/g \sim 0.01$ for $L/L_{\rm
Edd} = 0.01$, any line-driving below that level is insignificant for
these winds.  Thus, for initial densities $n_0 < 10^9~{\rm cm}^{-3}$,
line driving falls below the level of continuum driving and ceases to
be important.  For the highest density tested, $n_0 = 10^{11}~{\rm
cm}^{-3}$, line driving dominates continuum driving for all locations
in the wind.  (The variations in each acceleration curve as a function
of $s$ shown on this plot are chiefly due to variations in ionization
parameter: line-driving is high near the disk due to shielding and the
relatively high density, and rises towards the end of the streamline
due to the dropping flux levels at large distances.)

Since $n_0$ clearly has a great impact on the radiative acceleration,
how do such changes affect observables, such as the velocity?
Figure~\ref{vpVsDensity} shows how the variation in radiative
acceleration affects the poloidal velocity ($v_{\rm p}$) of the
outflow.  As the initial density and radiative acceleration increase,
the wind's velocity shows substantial variations from the pure
magnetocentrifugal model (which dominates the $n_0 = 10^{7}~{\rm
cm}^{-3}$ and $n_0 = 10^{8}~{\rm cm}^{-3}$ models).  In the case of
$n_0 = 10^{11}~{\rm cm}^{-3}$, where the greatest difference in
$v_{\rm p}$ is seen, the velocity increases by a factor of $\sim 2$ to
3 close to the disk.  Beyond the region close to the disk ($s/r_0 >
1$), however, magnetocentrifugal driving still dominates the final
velocities for these winds.  But the velocity differences near the
disk may be observationally important, especially if acceleration near
the base of the wind is the source of single-peaked emission lines
\citep[as in][]{MC97}.  If true, such emission lines may be critical
in testing the differences between wind models.

\subsection{Variations with Radiative Column}\label{radColumnVariations}

Having already tested the more obvious parameters of the initial
density and shield column density, we now turn to one of the most
crucial parameters for the efficiency of line-driving: the optical
depth in the lines.

\begin{figure}[h]
\begin{center}
\includegraphics[width=8cm]{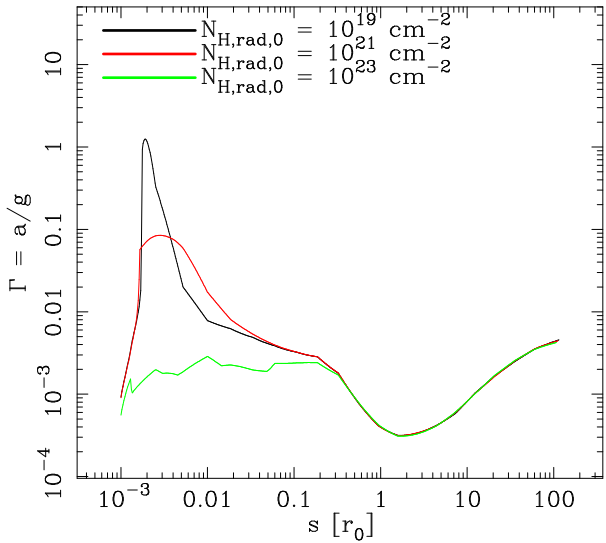}
\caption{Variations in line-driving as the radiative wind column is
  increased.  As the wind thickness increases, the optical depth
  within of the lines within the magnetocentrifugal wind increases and
  line-driving drops in strength. (Continuum-driving is approximately
  constant at $a/g \sim 0.01$.)
\label{Gamma_radColumnScan}}
\end{center}
\end{figure}

Under normal circumstances, where the gas velocity is sufficiently
low, or where the gas column is very low, the optical depths in the
lines are simply governed by the ionic columns themselves.  For large
accelerations or large columns, however, the optical depths are
dominated by the Sobolev length, which is defined as the distance over
which the relative velocity between atoms is equal to the thermal
width, so that a photon emitted by one atom could be absorbed by
another within that Sobolev length (see \S\ref{lineContAccel}).  With
the Eddington ratio in this magnetocentrifugal wind model ($L/L_{\rm
Edd} = 0.01$), both regimes can be important.  Depending on the
initial parameters prescribed, the wind column can be small enough
such that the column alone determines the opacity in the wind (instead
of the velocity) and therefore the amount of observed acceleration, so
that the Sobolev length is not important.  This is demonstrated in
Figure~\ref{Gamma_radColumnScan}, where the variation in line-driving
with the radiatively accelerated wind column is presented (the
shielding column is held constant).  For the larger columns, the
optical depth in the lines increases and the radiative acceleration
decreases.  For the smallest columns, very large radiative
acceleration is predicted due to the low opacity in the lines.  This
is critical for these models, for at significantly high column, the
wind would see no significant line driving (this is true for
magnetocentrifugal wind columns with $N_{\rm H,rad,0} \ga 10^{22}~{\rm
cm}^{-2}$).

\subsection{Modifying the Eddington Ratio}\label{eddRatioVariations}

\begin{figure*}[t]
\begin{center}
\includegraphics[width=15cm]{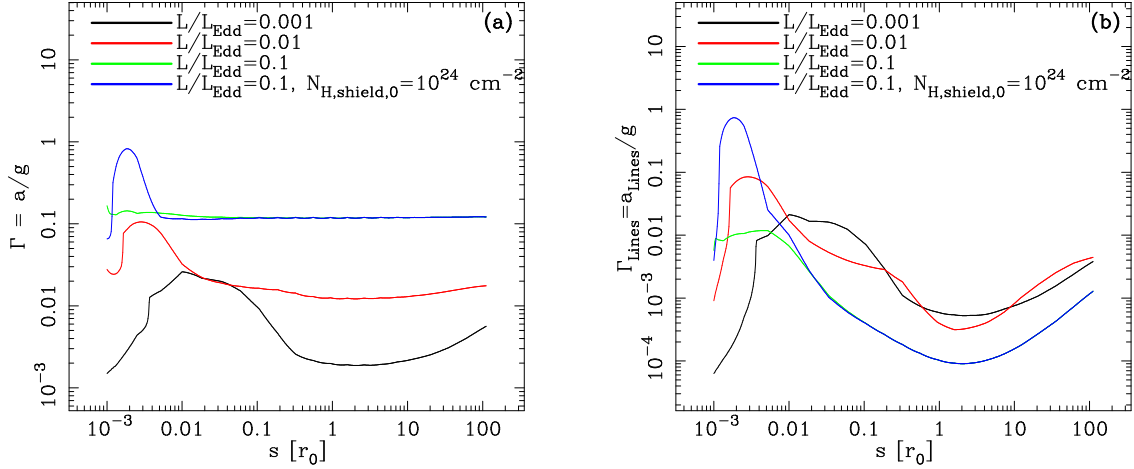}
\caption{Variation in both continuum and line driving (a) and line
  driving alone (b) when the Eddington ratio in the fiducial model is
  modified.  The first three models all have $N_{\rm H,shield,0} =
  10^{23}$~cm$^{-2}$, while the last model has $N_{\rm H,shield,0} =
  10^{24}$~cm$^{-2}$.  For lower Eddington ratios ($L/L_{\rm Edd} =
  0.001, 0.01$) the gas is in a relatively low ionization state for
  line-driving to be important, and the acceleration departs from the
  pure continuum acceleration (which sets the acceleration relative to
  gravity to about the Eddington ratio).  For a high Eddington ratio
  ($L/L_{\rm Edd} = 0.1$), the gas is overionized and only continuum
  acceleration is important; raising the shielding to a nearly
  Compton-thick $N_{\rm H,shield,0} = 10^{24}$~cm$^{-2}$ lowers the
  ionization state of the gas and allows much higher radiative
  acceleration.
\label{eddRatioScan}}
\end{center}
\end{figure*}

Of key importance to applications to AGN is understanding the
acceleration of outflows as a function of the Eddington ratio,
$L/L_{\rm Edd}$.  To investigate the impact of varying Eddington
ratios in our model, we present Figure~\ref{eddRatioScan}, which
displays the radiative acceleration (both the combined line and
continuum acceleration in panel \emph{a} as well as the line driving
in panel \emph{b}) in the fiducial model for three different Eddington
ratios: $L/L_{\rm Edd} = 0.001, 0.01,$ and $0.1$.  

The largest variation in Figure~\ref{eddRatioScan}a is the continuum
driving increasing linearly with the Eddington ratio.  As expected,
the continuum acceleration, relative to gravity, is roughly equal to
the Eddington ratio.  The line driving, on the other hand, can be seen
in both the deviations from the approximately constant continuum
acceleration in Figure~\ref{eddRatioScan}a and in
Figure~\ref{eddRatioScan}b.  

We begin examining this figure by concentrating on the first three
models in Figure~\ref{eddRatioScan}, which have $N_{\rm H,shield,0} =
10^{23}$~cm$^{-2}$.  In these models in Figure~\ref{eddRatioScan}a,
the increase in acceleration due to line-driving, relative to the
continuum-driving, decreases as the Eddington ratio increases.  This
can also be seen in Figure~\ref{eddRatioScan}b, where the line-driving
peaks near the disk ($s/r_0 \la 0.01$) for $L/L_{\rm Edd} = 0.01$, but
decreases for Eddington ratios an order of magnitude larger (where the
gas is overionized) and an order of magnitude smaller (where the
radiation field doesn't have the momentum to accelerate the wind as
strongly as at $L/L_{\rm Edd} = 0.01$).

Now we turn to the fourth model in Figure~\ref{eddRatioScan}, where we
keep $L/L_{\rm Edd} = 0.1$ but increase the shielding level to $N_{\rm
H,shield,0} = 10^{24}~$cm$^{-2}$.  This model shows the importance of
shielding gas for $L/L_{\rm Edd} = 0.1$.  The increase in $L/L_{\rm
Edd}$ and the increase in shielding relative to the fiducial model
allow for increased line radiative acceleration that jumps almost two
orders of magnitude in strength near the base of the wind.

\section{Results}\label{results}

If magnetocentrifugal winds power outflows in AGNs, they must
certainly be affected by the intense radiation field they experience;
in turn, such winds will also influence the efficiency of radiative
acceleration.  This paper has explored the radiative transfer through
these magnetocentrifugal winds, and how they both are affected by and
affect radiative acceleration of outflows from AGNs.  The model has
been used to explore the detailed dynamics and ionization of a
fiducial magnetocentrifugal disk wind, showing the inter-relation
between shielding column, initial number density, outflow velocity,
Eddington ratio, and acceleration.

As a result of this study, these models have shown:

\begin{enumerate} 

\item A magnetocentrifugal outflow, acting as a ``shield'', can
improve the efficiency of line-driving by factors of approximately two
to three $N_{\rm H,shield,0} = 10^{23}~$cm$^{-2}$
(\S\ref{shieldColumnVariations}) and by up to almost two orders of
magnitude for $N_{\rm H,shield,0} = 10^{24}~$cm$^{-2}$
(\S\ref{eddRatioVariations}).  A magnetocentrifugal wind has the
advantage that it can be accelerated without regard to the ionization
state, whereas radiatively-driven winds must have a low ionization
parameter in order for a critical abundance of atomic lines to be
present.  Therefore, magnetocentrifugal winds could play an important
role in acting as a radiation shield and allow large radiative
accelerations.  It may also be possible that pressure differences
(MCGV95) or disk photons \citep{PK04} may help ``lift'' the shield;
neither of those effects are considered here.  Later work with this
model will include the effect of disk-emitted photons.

\item The efficiency of line-driving is strongly dependent on the
density at the base of the wind (\S\ref{nInitVariations}).  This is
due to the very critical dependence of line acceleration on the
ionization parameter.  The lower the ionization state, the more lines
exist to aid in the momentum transfer from outward-streaming photons.
The density at the base of the disk is therefore crucial to setting
to line-driving within these magnetocentrifugal models.

\item Small columns ($N_{\rm H,rad,0} \la 10^{21}~{\rm cm}^{-2}$)
within magnetocentrifugal winds can be significantly accelerated by
line-driving (\S\ref{radColumnVariations}).  This point demonstrates
the importance of ``non-Sobolev'' effects; i.e., that at low columns,
the optical depths in the lines drop below the opacity given by the
Sobolev length, and at such low columns, the radiative acceleration
can be underestimated by the simple Sobolev approximation.
\end{enumerate}

In addition, by examining the fiducial model and the above cases where
model parameters are varied, these solutions have displayed the
importance of considering the detailed interaction between the
dynamics and photoionization in AGN outflows.  Calculations of the
ionization parameter along the flow and in the variation of shielding
and optical depth along the flow are central issues to modeling these
winds.  The issues outlined above are a few of the dependences that
arise from such modeling, and may indeed (by the variation in
acceleration and therefore velocity along the streamlines) lead to
tests to observationally determine the physics of wind launching in
AGNs.

Thus, while this model has been developed to help address the above
questions about shielding and the affects of radiative driving on
magnetocentrifugal winds, the solutions available are not limited to
the above, examined cases.  Future papers will study further the
variation of radiative acceleration on model parameters such as the
SED, atomic line lists used to calculate the acceleration, initial
densities, Eddington ratios, and other parameters of the model.  The
model can also be used to explore the absorption and emission features
from such a wind, as well as, for instance, the possible role of
``clouds'' within a continuous wind \citep{EKA02}.  In addition, this
model is in no way constrained to only study AGNs.  The same basic
physical framework could also be employed to study winds from
accretion disks surrounding young stellar objects or cataclysmic
variables.

\section{Acknowledgments}
I gratefully acknowledge my advisor, Arieh K\"onigl, for many useful
conversations throughout the course of this project.  In addition,
many thanks to Gary Ferland and his collaborators for developing
Cloudy, making it freely available, and supporting it.  The referee's
comments were very helpful, and those comments helped improve the
paper.  Thanks also to David Ballantyne, Pat Hall, Lewis Hobbs, John
Kartje, Ruben Krasnopolsky, Bob Rosner, Nektarios Vlahakis, and Don
York for valuable comments. This work would not have been possible
without the generous support of NASA's ATP program, in this case via
grant NAG5-9063, and the support of the Natural Sciences and
Engineering Research Council of Canada.  This research has made use of
NASA's Astrophysics Data System.

\appendix

\section{Appendix A: Derivation of the Self-Similar Centrifugal Wind Equations}\label{selfSimWindEqns}

In this appendix, a rederivation of the system of self-similar wind
equations for the magnetocentrifugal wind is presented.  The equations
utilized in this calculation advance upon those presented in BP82 and
KK94: the wind not only has an arbitrary density power-law index, $b$,
as in KK94, but energy conservation is not required.  Since the
radiation field continually inputs energy into the outflow, this is an
important modification that was not fully considered in the derivation
presented in KK94.

First, a stationary, axisymmetric, ideal, cold MHD flow in cylindrical
coordinates $(r,\phi,z)$ is assumed.  The equations are based on both
the radial and vertical momentum equations:
\begin{eqnarray}
v_r \frac{\partial v_r}{\partial r} + v_z \frac{\partial v_r}{\partial
z} - \frac{v_{\phi}^2}{r} & = & - \rho \frac{\partial \Phi}{\partial r} - 
\frac{B_z}{4 \pi} \left( \frac{\partial B_z}{\partial r} -
\frac{\partial B_r}{\partial z} \right) - \frac{B_{\phi}}{4 \pi r}
\frac{\partial (r B_{\phi})}{\partial r} \\
\rho (\mathbf{v} \cdot \mathbf{\nabla}) v_z & = & -\rho \frac{\partial
\Phi}{\partial z} - \frac{1}{8 \pi} \frac{\partial B^2}{\partial z} +
\frac{1}{4 \pi} (\mathbf{B} \cdot \mathbf{\nabla}) B_z, 
\end{eqnarray}
where $\mathbf{v}$ is the fluid velocity and $\mathbf{B}$ is the
magnetic field.  The thermal term is neglected in the limit that
thermal affects are much less important than magnetocentrifugal and
radiative-driving effects.  $\Phi$ is the effective gravitational
potential, defined as
\begin{eqnarray}
\Phi = - [1 - \Gamma(\theta)] \frac{GM_{\bullet}}{(r^2 + z^2)^{1/2}}\label{gammaInSelfSim} ,
\end{eqnarray}
where $M_{\bullet}$ is the mass of the central black hole, and
$\Gamma(\theta)$ gives the local radiative-to-gravitational radial
acceleration (see eq.~[\ref{gammaDef}]).

These equations are solved by first stipulating mass conservation,
\begin{eqnarray}
\mathbf{\nabla} \cdot (\rho \mathbf{v}) = 0, 
\end{eqnarray}
and then relating the flow velocity to the magnetic field via
\begin{eqnarray}
\mathbf{v}(\mathbf{r}) = \frac{k \mathbf{B}(\mathbf{r})}{4 \pi
\rho(\mathbf{r})} + \mathbf{\omega}(\mathbf{r}) \times \mathbf{r}, 
\end{eqnarray} 
\citep[e.g., ][]{Chandra56, Mestel61}, where $k/4\pi$ is the ratio of
mass flux to magnetic flux, and $\mathbf{\omega}(\mathbf{r})$ and
$\rho(\mathbf{r})$ are the field angular velocity and gas mass density
of the flow, respectively.  Both $\omega$ and $k$ are constant along
magnetic fieldlines.  In addition, while the specific energy is not
constant, the total specific angular momentum
\begin{eqnarray}
l = r v_{\phi} - \frac{r B_{\phi}}{k}
\end{eqnarray}
is conserved.

Self-similarity is then imposed on this system by specifying
\begin{eqnarray}
\mathbf{r} & = & [r_0 \xi(\chi), \phi, r_0 \chi], \\
\mathbf{v} & = & [\xi'(\chi) f(\chi), g(\chi), f(\chi)]v_{k,0},
\end{eqnarray}
where $v_{\rm k,0}$ is the Keplerian speed at the base of the outflow,
$v_{\rm k,0} = (GM_{\bullet}/r_0)^{1/2}$, and the prime indicates
differentiation with respect to $\chi$.  At the same time, the above
constants are re-expressed in dimensionless form:
\begin{eqnarray}
\lambda & \equiv & \frac{l}{(GM_{\bullet} r_0)^{1/2}}, \\
\kappa & \equiv & \frac{k (1+{\xi'}_0^2)^{1/2}}{B_{p,0}} v_{k,0},
\end{eqnarray}
where $B_{p,0}$ is the poloidal magnetic field strength at the base of
the wind.

As in KK94, a general power-law scaling of the density and magnetic
field along the disk's surface is defined:
\begin{eqnarray}
\rho_0 & \propto & r_0^{-b}, \\
B_0 & \propto & r_0^{-(b+1)/2}.
\end{eqnarray}

With this self-similar specification, the radial and vertical momentum
equations become, after some simplification:
\begin{eqnarray}
\frac{f \xi' m'}{\kappa \xi J} - \frac{f^2 \xi'}{\xi J} +
\xi'' f^2 - \frac{(\lambda m - \xi^2)^2}{\xi^3 (m-1)^2} & = & 
- \xi [1 - \Gamma(\theta)] S^3 - \frac{f}{\kappa \xi J^2} \left( 
\frac{-(1+\xi'^2)(b+1)}{2} + \right. \nonumber \\ 
& & \left. \frac{(\chi + \xi \xi')\xi'}{\xi} -
\frac{\xi''}{J S^2} \right) - \frac{\kappa f}{\xi} \frac{(\lambda -
\xi^2)}{(m-1)} \nonumber \\ 
& & \left[ \frac{(\lambda - \xi^2)}{(m-1)}\frac{(-b+1)}{2}
  + \right. \nonumber \\
& & \left. \chi \left( \frac{2 \xi \xi'}{(m-1)} + \frac{(\lambda - \xi^2)m'}
{(m-1)^2} \right) \right], \label{radMomEq} \\
\frac{f}{\kappa \xi J} (m' - f \kappa \xi' J + f \kappa \xi
\chi \xi'')  & = &
 -[1 - \Gamma(\theta)]\chi S^3 + \frac{f \xi'}{\kappa \xi J^2}
 \left( \frac{-(1+\xi'^2)(b+1)}{2} + \nonumber \right. \\
& & \left. \frac{(\chi + \xi \xi')\xi'}{\xi} -
\frac{\xi''(\chi^2 + \xi^2)}{J} \right) - \nonumber \\
& & \xi' \kappa f (\lambda - \xi^2)
\left(\frac{(b+1)(\lambda-\xi^2) - 2(\lambda+\xi^2)}{2 \xi (m-1)^2} \right) + 
\nonumber \\
& & \frac{(\lambda - \xi^2)^2 m' \kappa f}{(m-1)^3}, \label{vertMomEq}
\end{eqnarray}
where
\begin{eqnarray}
m & \equiv & \frac{4 \pi \rho v_p^2}{B_p^2} = \kappa \xi f J = {\rm
square~of~poloidal~Alfv\acute{e}n~Mach~number}, \\
\kappa & \equiv & \frac{k (1 - {\xi'}_0^2)^{\frac{1}{2}} v_{k,0}}{B_{p,0}}  =
{\rm dimensionless~ratio~of~mass~flux~to~magnetic~flux},\\
\lambda & \equiv & \frac{l}{(G M r_0)^{\frac{1}{2}}} = {\rm normalized~angular~momentum},\\
J & \equiv & \xi - \chi \xi', \\
S & \equiv & 1/\sqrt{\xi^2 + \chi^2}. \\
\end{eqnarray}
The two equations~(\ref{radMomEq}) and (\ref{vertMomEq}) define the
differential equations for $m'$ and $\xi''$, which are, respectively,
the spatial gradient in the poloidal Alfv\'en mach number (gradient
with respect to height, $\chi$) and the (cylindrical) radial velocity
gradient (again with respect to $\chi$).

One can see from close inspection of the above equations that many of
the terms have a denominator of $(m - 1)$, showing that when the gas
crosses the Alfv\'en point (where $m = 1$), the equations become
singular.  Rewriting and solving the $m'$ equation for the value of
$m'$ at the Alfv\'en singular point:
\begin{eqnarray}
m'_{A} & = & 2 \xi J [-8 \chi \kappa^2 \lambda m' \xi' J^3 + 4 (1+b)
\kappa^2 \lambda \xi'^2 J^2 (\chi + \xi \xi') + m'^2 (\chi + \xi \xi')
(-2 \kappa^2 \lambda^2 S + \nonumber \\
& & (1+b) + 2 \kappa^2 \lambda^3 - 4 \chi
\kappa^2 \lambda^{\frac{3}{2}}(\lambda - S) \xi' + ( (1+b) + 2 \chi^2
\kappa^2 \lambda (\lambda - S))\xi'^2 + 2 \kappa^2 \lambda
\Gamma(\theta) J^2)] / \nonumber \\
& & \left[ 4 \xi J \left( \frac{4 \kappa^2 \lambda \xi'^2
J^2}{S^2} + m'^2 (\chi + \xi \xi')^2 \right) \right]. \label{mpAeqn}
\end{eqnarray}
This constraint is used to start the integral at the Alfv\'en point
with the value of $m'_A$ given by equation~(\ref{mpAeqn}).

As covered in the main text, these equations are solved using a
``shooting algorithm,'' integrating from both the Alfv\'en point and
the disk surface towards an intermediate point.  Matching the
integrals of three first-order equations (given by the first-order
equation for $m'$ and the second-order equation for $\xi''$ in
eqs.~\ref{radMomEq} and \ref{vertMomEq}) at the common point allows us
to solve for the three free parameters in the system: $\xi'_0$,
$\xi'_A$, and $\chi_A$.


\begin{thebibliography}{}
\bibitem[Abbott(1980)]{Abbott80}Abbott, D.C. 1980, ApJ, 242, 1183
\bibitem[Antonucci(1993)]{Ant93}Antonucci, R. 1993, ARA\&A, 31, 473
\bibitem[Arav et al.(1994)Arav, Li, \& Begelman]{Arav94}Arav, N., Li, Z., \&
Begelman, M.C. 1994, ApJ, 432, 62
\bibitem[Arav et al.(1995)]{Arav95}Arav, N., Korista, K.T., Barlow,
  T.A., Begelman, M.C. 1995, Nature, 376, 576
\bibitem[Arav(1996)]{Arav96} Arav. N. 1996, ApJ, 465, 617
\bibitem[Arav et al.(1998)]{Arav98} Arav, N., Barlow, T.A., Laor, A.,
Sargent, W.L.W, \& Blandford, R.D. 1998, MNRAS, 297, 990
\bibitem[Arnaud \& Raymond(1992)]{AR92}Arnaud, M. \& Raymond, J. 1992,
  ApJ, 398, 394
\bibitem[Baldwin et al.(1991)]{Baldwin91}Baldwin, J., Ferland, G.J.,
Martin, P.G., Corbin, M., Cota, S., Peterson, B.M., \& Slettebak, A.
1991, ApJ, 374, 580
\bibitem[Baldwin et al.(1995)]{Baldwin95}Baldwin, J., Ferland, G.J.,
Korista, K., Verner, D. 1995, ApJ, 455, 119
\bibitem[Balsara \& Krolik(1993)]{BK93}Balsara, D. \& Krolik,
  J.H. 1993, ApJ, 402, 109
\bibitem[Blandford \& K\"onigl(1979)]{BK79} Blandford, R.D., \&
K\"onigl, A. 1979, Astrophys. Lett., 20, 15
\bibitem[Blandford \& Payne(1982)]{BP82}Blandford, R.D., \& Payne,
D.G. 1982, MNRAS, 199, 883 (BP82)
\bibitem[Blandford(2001)]{Blandford2001}Blandford, R.D. 2001, in ASP
Conf. Ser. 224, Probing the Physics of Active Galactic Nuclei by
Multiwavelength Monitoring, ed. B. M. Peterson, R. S. Polidan \&
R. W. Pogge (San Francisco: ASP), 499
\bibitem[Bottorff et al.(1997)]{Bot97}Bottorff, M., Korista, K.T.,
Shlosman, I., \& Blandford, R.D. 1997, ApJ, 479, 200
\bibitem[Bottorff, Korista, \& Shlosman(2000)]{BKS00}Bottorff, M., Korista,
  K.T., \& Shlosman, I. 2000, ApJ, 537, 134
\bibitem[Bottorff \& Ferland(2001)]{Bot01}Bottorff, M., \& Ferland,
G. 2001, ApJ, 549, 118
\bibitem[Cassinelli(1979)]{C79}Cassinelli, J. P. 1979, ARA\&A, 17, 275
\bibitem[Castor et al.(1975)Castor, Abbott, \& Klein]{CAK75}Castor, J.I., Abbott,
D.C., \& Klein, R.I. 1976, ApJ, 195, 157
\bibitem[Chandrasekhar(1956)]{Chandra56}Chandrasekhar, S. 1956, ApJ,
124, 232
\bibitem[Chelouche \& Netzer(2001)]{CN01} Chelouche, D.,
Netzer, H. 2001, MNRAS, 326, 916
\bibitem[Chelouche \& Netzer(2003a)]{CN03a} Chelouche, D., Netzer,
H. 2003, MNRAS, 344, 223
\bibitem[Chelouche \& Netzer(2003b)]{CN03b} Chelouche, D., Netzer,
H. 2003, MNRAS, 344, 233
\bibitem[Chiang \& Blaes(2003)]{CB03}Chiang, J. \& Blaes, O. 2003,
  ApJ, 586, 97
\bibitem[Contopoulos \& Lovelace(1994)]{CL94}Contopoulos, J., \&
Lovelace, R.V.E. 1994, ApJ, 429, 139
\bibitem[Crenshaw et al.(1999)]{C99}Crenshaw, D.M., Kraemer, S.B.,
  Boggess, A., Maran, S.P., Mushotzky, R.F., Wu, C. 1999, ApJ, 516,
  750
\bibitem[Crenshaw, Kraemer, \& George(2003)]{CKG03}Crenshaw, D.M.,
  Kraemer, S.B., George, I.M. 2003, ARAA, 41, 117
\bibitem[de Kool \& Begelman(1995)]{dKB95}de Kool, M., \& Begelman,
M.C. 1995, ApJ, 455, 448
\bibitem[Deluit(2004)]{Deluit04}Deluit, S.J. 2004, A\&A, 415, 39
\bibitem[Draine \& Lee(1984)]{DL84}Draine B.T., \& Lee, H.M. 1984, ApJ, 285, 89
\bibitem[Drew \& Boksenberg(1984)]{DB84}Drew, J.E., \& Boksenberg,
  A. 1984, MNRAS, 211, 813
\bibitem[Elvis(2000)]{Elvis00}Elvis, M. 2000, ApJ, 545, 63
\bibitem[Elvis, Risaliti, \& Zamorani(2002)]{Elvis02}Elvis, M.,
  Risaliti, G., \& Zamorani, G. 2002, ApJ, 565, L75
\bibitem[Emmering et al.(1992)Emmering, Blandford, \&
Shlosman]{EBS92}Emmering, R.T., Blandford, R.D., \& Shlosman,
I. 1992, ApJ, 385, 460
\bibitem[Everett et al.(2002)Everett, K\"onigl, \& Arav]{EKA02}Everett, J.E.,
K\"onigl, A., Arav, N. 2002, ApJ, 569, 671
\bibitem[Fath(1909)]{Fath09}Fath, E.A. 1909, Lick Obs. Bull., 149, 71
\bibitem[Feldmeier \& Shlosman(1999)]{FS99}Feldmeier, A. \& Shlosman,
  I. 1999, ApJ, 526, 344
\bibitem[Ferland et al.(1998)]{F98}Ferland, G.J., Korista,
K.T., Verner, D.A., Ferguson, J.W., Kingdon, J.B., \& Verner, E.M. 1998,
PASP, 110, 761
\bibitem[Ferreira(2003)]{F03}Ferreira, J. 2003 in Star Formation and
  the Physics of Young Stars, eds. J. Bouvier and J.-P. Zahn (Les
  Ulis: EDP Sciences), 229 (astro-ph/0311621)
\bibitem[Friend \& MacGregor(1984)]{FM84}Friend,  D.B. \& MacGregor,
  K.B. 1984, ApJ, 282, 591
\bibitem[Fromerth \& Melia(2001)]{FM01}Fromerth, M.J \& Melia,
  F. 2001, ApJ, 549, 205
\bibitem[Ganguly et al.(2001)]{Ganguly01}Ganguly, R., Bond, N.A.,
Charlton, J.C., Eracleous, M., Brandt, W.N., \& Churchill, C.W. 2001,
ApJ, 549, 133
\bibitem[Gregori et al.(2000)]{Gregori00}Gregori, G., Miniati, F.,
  Ryu, D., \& Jones, T.W. 2000, ApJ, 543, 775
\bibitem[Hamann et al.(2001)]{Ham01}Hamann, F.W., Barlow, T.A.,
Chaffee, F.C., Foltz, C.B., \& Weymann, R.J. 2001, ApJ, 550, 142
\bibitem[Hartley et al.(2002)]{Hartley02}Hartley, L.E., Drew, J.E.,
  Long, K.S., Knigge, C., Proga, D. 2002, MNRAS, 332, 127
\bibitem[Kartje(1995)]{K95}Kartje, J.F. 1995, ApJ, 452, 565
\bibitem[Kartje et al.(1999)Kartje, K\"onigl, \&
Elitzur]{KKE99} Kartje, J.F., K\"onigl, A., \& Elitzur, M. 1999, ApJ,
513, 180
\bibitem[Krolik(1999)]{Krolik99}Krolik, J.H. 1999, Active Galactic
Nuclei: From the Central Black Hole to the Galactic Environment
(Princeton: Princeton University Press)
\bibitem[Krolik \& Kriss(2001)]{KK01}Krolik, J.H. \& Kriss, G.A. 2001,
  ApJ, 561, 684
\bibitem[K\"onigl \& Kartje(1994)]{KK94}K\"onigl, A., \& Kartje,
J.F. 1994, ApJ, 434, 446 (KK94)
\bibitem[K\"onigl \& Pudritz(2000)]{KP00}K\"onigl, A., \& Pudritz,
R.E. 2000 in Protostars \& Planets IV, ed. V. Mannings, A. P. Boss,
S. S. Russell (Tucson: University of Arizona Press), 759
\bibitem[Kraemer et al.(2002)]{Kraemer02}Kraemer, S.B., Crenshaw,
  D.M., George, I.M., Netzer, H., Turner, T.J., Gabel, J.R. 2002, ApJ,
  577, 98
\bibitem[Krasnopolsky, Li, \& Blandford(1999)]{K99}Krasnopolsky, R.,
  Li, Z.-Y.,Blandford, R. 1999, ApJ, 526, 631
\bibitem[Kuncic, Celotti, \& Rees(1997)]{Kuncic97}Kuncic, Z., Celotti,
  A., Rees, M.J. 1997, MNRAS, 284, 717
\bibitem[Laor \& Brandt(2002)]{LB02} Laor, A., \& Brandt,
W.N. 2002, ApJ, 569, 641
\bibitem[Martin \& Rouleau(1991)]{MR91}Martin, P.G., \& Rouleau, F. 1991,
in Extreme Ultraviolet Astronomy, eds, Malina, R.F., Bowyer S., 
Pergamon Press, Oxford, 341
\bibitem[Mathews \& Ferland(1987)]{MF87}Mathews, W.G., \& Ferland,
G.J. 1987, ApJ, 323, 456 (MF87)
\bibitem[Mathis, Rumpl, \& Nordsieck(1977)]{MRN77}Mathis, J.S., Rumpl,
W., \& Nordsieck, K.H. 1977, ApJ, 217, 425 
\bibitem[Meier et al.(1997)]{Meier97}Meier, D.L., Edgington, S.,
Godon, P., Payne, D.G., Lind, K.R. 1997, Nature, 388, 350
\bibitem[Mestel(1961)]{Mestel61}Mestel, L. 1961, MNRAS, 122, 473
\bibitem[Mihalas \& Weibel-Mihalas(1999)]{MW99}Mihalas, D. \&
  Weibel-Mihalas, B. 1999, Foundations of Radiation Hydrodynamics,
  (Mineola:Dover Publications, Inc.)
\bibitem[Murray \& Chiang(1997)]{MC97}Murray, N., \& Chiang, J. 1997,
ApJ, 474, 91
\bibitem[Murray et al.(1995)]{MCGV95}Murray, N., Chiang, J.,
Grossman, S.A., \& Voit, G.M. 1995, ApJ,451, 498 (MCGV95)
\bibitem[Netzer(1990)]{Netzer90}Netzer, H. 1990, in Active Galactic Nuclei,
Saas-Fee Advanced Course 20 (Berlin: Springer-Verlag), 57
\bibitem[Ouyed \& Pudritz(1997)]{OP97}Ouyed, R., Pudritz, R.E. 1997,
  ApJ, 482, 712
\bibitem[Ouyed et al.(1997)]{Ouyed97}Ouyed, R., Pudritz, R.E., Stone,
  J.M. 1997, Nature, 385, 409
\bibitem[Pereyra et al.(2004)]{Pereyra04}Pereyra, N.A., Owocki, S.P.,
  Hillier, D.J., Turnshek, D.A. 2004, ApJ, 608, 454
\bibitem[Peterson(1997)]{Peterson97}Peterson, B.M. 1997, An
Introduction to Active Galactic Nuclei, (New York:Cambridge University
Press)
\bibitem[Powell(1970)]{Powell70}Powell, M.J.D. 1970, in Numerical
Methods for Nonlinear Algebraic Equations, ed, P. Rabinowitz (New
York: Gordon and Breach).
\bibitem[Proga(1999)]{P99}Proga, D. 1999, MNRAS, 304, 938
\bibitem[Proga(2000)]{P00}Proga, D. 2000, ApJ, 538, 684
\bibitem[Proga(2003)]{P03}Proga, D. 2003, ApJ, 585, 406
\bibitem[Proga, Stone \& Drew(1998)]{PSD98}Proga, D., Stone, J.M.,
  Drew, J.E. 1998, MNRAS, 295, 595
\bibitem[Proga, Stone, \& Kallman(2000)]{PSK00}Proga, D., Stone, J.M.,
\& Kallman, T.R. 2000, ApJ, 543, 686, PSK00
\bibitem[Proga \& Kallman(2004)]{PK04}Proga, D., \& Kallman, T.R,
  2004, accepted to ApJ (astro-ph/0408293) 
\bibitem[Rees(1987)]{Rees87}Rees, M.J. 1987, MNRAS, 228, 47P
\bibitem[Reichard et al.(2003)]{Reichard03}Reichard, T.A., Richards,
  G,T., Hall, P.B., Schneider, D.P., Vanden Berk, D,E., Fan, X., York,
  D.G.; Knapp, G.R.; Brinkmann, J. 2003, AJ, 126, 2594
\bibitem[Risaliti \& Elvis(2004)]{RE04}Risaliti, G \& Elvis, M. 2004,
  in Supermassive Black Holes in the Distant Universe,
  ed. A. J. Barger (Boston: Kluwer)
\bibitem[Romanova et al.(1997)]{Romanova97} Romanova, M.M., Ustyugova,
G.V., Koldoba, A.V., Chechetkin, V.M., Lovelace, R.V.E. 1997, ApJ,
482, 708
\bibitem[Safier(1993)]{S93}Safier, P. 1993, ApJ, 408, 115
\bibitem[Sobolev(1958)]{Sobolev58}Sobolev, V.V. 1958 in Theoretical
Astrophysics, ed. V.A. Ambartsumian (London: Pergamon)
\bibitem[Spruit(1996)]{S96}Spruit, H.C. 1996 in Physical processes in
  Binary Stars, eds. R.A.M.J. Wijers, M.B. Davies \& C.A. Tout,
  (Dordrecht: Kluwer) (astro-ph/9602022)
\bibitem[Steenbrugge et al.(2003)]{Steenbrugge03}Steenbruge, K.C.,
Kaastra, J.S., de Vries, C.P., Edelson, R. 2003, A\&A in press
(astro-ph/0302493)
\bibitem[Stevens \& Kallman(1990)]{SK90}Stevens, I.R, \& Kallman,
T.R. 1990, ApJ, 365, 321
\bibitem[Tielens et al.(1994)]{Tielens94}Tielens, A.G.G.M., McKee,
C.F., Seab, C.G., Hollenbach, D.J. 1994, ApJ, 431, 321
\bibitem[Tran(2003)]{Tran03}Tran, H.D. 2003, ApJ, 583, 632
\bibitem[Urry \& Padovani(1995)]{UP95}Urry, C.M., \& Padovani, P. 1995,
PASP, 107, 803
\bibitem[Ustyugova et al.(1995)]{Ustyugova95}Ustyugova, G.V., Koldoba,
  A.V., Romanova, M.M., Chechetkin, V.M., \& Lovelace, R.V.E. 1995,
  ApJ, 439, L39
\bibitem[Verner \& Ferland(1996)]{VF96}Verner, D.A., \& Ferland,
  G.J. 1996, ApJS, 103, 467
\bibitem[Verner et al.(1996)]{Verner96}Verner, D.A., Verner, E.M., \&
Ferland, G.J. 1996, Atomic Data Nucl. Data Tables, 64, 1
\bibitem[Vitello \& Shlosman(1988)]{VS88}Vitello, P.A.J, \& Shlosman,
  I. 1988, ApJ, 327, 680
\bibitem[Vlahakis et al.(2000)]{V00} Vlahakis, N., Tsinganos, K.,
  Sauty, C., Trussoni, E., 2000, MNRAS, 318, 417
\bibitem[Wardle \& K\"onigl(1993)]{WK93}Wardle, M., \& K\"onigl,
A. 1993, ApJ, 410, 218
\bibitem[Weber \& Davis(1967)]{WD67}Weber, E.J. \& Davis, L. Jr. 1967,
  ApJ, 148, 217
\bibitem[Weymann et al.(1991)]{Weymann91}Weymann, R.J., Morris,
  S.L., Foltz, C.B., Hewett, P.C. 1991, ApJ, 373, 23
\end{thebibliography}
\end{document}